\documentclass{aa}  
\usepackage{graphicx}
\usepackage{txfonts}

\usepackage[switch]{lineno} 
\usepackage{amsmath}    
\usepackage{amssymb}    
\usepackage{cases}  
\usepackage{multirow} 
\usepackage{xcolor} 

\usepackage{xfrac}
\usepackage{etoolbox}
\makeatletter
\patchcmd\@combinedblfloats{\box\@outputbox}{\unvbox\@outputbox}{}{%
   \errmessage{\noexpand\@combinedblfloats could not be patched}%
}%
 \makeatother
 
\DeclareRobustCommand*{\drv}{\mathop{}\!\mathrm{d}}
\DeclareRobustCommand*{\kms}{km\,s$^{-1}$}

\begin{document}

   \title{Solar wind charge exchange in cometary atmospheres}

   \subtitle{III. Results from the Rosetta mission to comet 67P/Churyumov-Gerasimenko}

   \author{Cyril Simon Wedlund\inst{1}
\and Etienne Behar\inst{2,3}
\and Hans Nilsson\inst{2,3}
\and Markku Alho\inst{4}
\and Esa Kallio\inst{4}
\and Herbert Gunell\inst{5,6}
\and Dennis Bodewits\inst{7}
\and Kevin Heritier\inst{8}
\and Marina Galand\inst{8}
\and Arnaud Beth\inst{8}
\and Martin Rubin\inst{9}
\and Kathrin Altwegg\inst{9}
\and Martin Volwerk\inst{10}
\and Guillaume Gronoff\inst{11,12}
\and Ronnie Hoekstra\inst{13}
}

   \institute{Department of Physics, University of Oslo, P.O. Box 1048 Blindern, N-0316 Oslo, Norway\\
              \email{cyril.simon.wedlund@gmail.com}
         \and
             Swedish Institute of Space Physics, P.O. Box 812, SE-981 28 Kiruna, Sweden
         \and
             Lule\aa{} University of Technology, Department of Computer Science, Electrical and Space Engineering, Kiruna,  SE-981 28, Sweden
        \and
            Department of Electronics and Nanoengineering, School of Electrical Engineering, Aalto University, P.O. Box 15500, 00076 Aalto, Finland
        \and
            Royal Belgian Institute for Space Aeronomy, Avenue Circulaire 3, B-1180 Brussels, Belgium
            \and
            Department of Physics, Ume\aa{} University, 901 87 Ume\aa{}, Sweden
        \and 
            Physics Department, Auburn University, Auburn, AL 36849, USA
        \and 
            Department of Physics, Imperial College London, Prince Consort Road, London SW7 2AZ, United Kingdom
        \and
            Space Research and Planetary Sciences, University of Bern, 3012 Bern, Switzerland
        \and 
            Space Research Institute, Austrian Academy of Sciences, Schmiedlstra{\ss}e 6, 8042 Graz, Austria
        \and
            Science directorate, Chemistry \& Dynamics branch, NASA Langley Research Center, Hampton, VA 23666 Virginia, USA
        \and
            SSAI, Hampton, VA 23666 Virginia, USA
        \and 
            Zernike Institute for Advanced Materials, University of Groningen, Nijenborgh 4, 9747 AG, Groningen, The Netherlands
        }

   \date{\today}

 
  \abstract
   {Solar wind charge-changing reactions are of paramount importance to the physico-chemistry of the atmosphere of a comet. The ESA/\emph{Rosetta} mission to comet 67P/Churyumov-Gerasimenko (67P) provides a unique opportunity to study charge-changing processes in situ.}
   {To understand the role of these reactions in the evolution of the solar wind plasma and interpret the complex in situ measurements made by \emph{Rosetta}, numerical or analytical models are necessary.}
   {We used an extended analytical formalism describing solar wind charge-changing processes at comets along solar wind streamlines. The model is driven by solar wind ion measurements from the Rosetta Plasma Consortium-Ion Composition Analyser (RPC-ICA) and neutral density observations from the Rosetta Spectrometer for Ion and Neutral Analysis-Comet Pressure Sensor (ROSINA-COPS), as well as by charge-changing cross sections of hydrogen and helium particles in a water gas.}
   {A mission-wide overview of charge-changing efficiencies at comet 67P is presented. Electron capture cross sections dominate and favor the production of He and H energetic neutral atoms (ENAs), with fluxes expected to rival those of H$^{+}$ and He$^{2+}$ ions.}
   {Neutral outgassing rates are retrieved from local RPC-ICA flux measurements and match ROSINA estimates very well throughout the mission. From the model, we find that solar wind charge exchange is unable to fully explain the magnitude of the sharp drop in solar wind ion fluxes observed by \emph{Rosetta} for heliocentric distances below $2.5$\,AU. This is likely because the model does not take the relative ion dynamics into account and to a lesser extent because it ignores the formation of bow-shock-like structures upstream of the nucleus. 
   This work also shows that the ionization by solar extreme-ultraviolet radiation and energetic electrons dominates the source of cometary ions, although solar wind contributions may be significant during isolated events.
   }


   \maketitle
%

\section{Introduction}
Between August 2014 and its controlled end-of-mission crash on 30 September 2016, the European Space Agency mission \emph{Rosetta} accompanied comet 67P/Churyumov-Gerasimenko (67P) on its journey toward the inner solar system, through perihelion at $1.24$~astronomical units (AU), and back out again \citep{Jones2017,Taylor2017}. During this period, \emph{Rosetta} monitored in situ the neutral and plasma environment of the comet and its interaction with the solar wind, using dedicated instruments \citep{Glassmeier2017}. 

A relevant process in the collisional interaction between a neutral environment and the solar wind is charge exchange \citep[see][]{Dennerl2010}. In the following, "charge exchange" and  "charge transfer" are used interchangeably to denote electron capture reactions only, whereas charge-changing collisions encompass both electron capture and stripping reactions. In cometary or planetary atmospheres, charge-exchange reactions, such as single or multiple electron captures, result in the creation of slow cometary ions, arising from collisions between fast solar wind ions and the slow-moving neutral gas. In the process, solar wind ions may become excited and emit radiation in the form of soft X-rays \citep[see][and references therein]{Cravens2004}. For a solar wind made of X$^{i+}$ ions impacting a neutral gas species M, the electron capture of $q$ electrons is
\begin{align}
        \textnormal{X}^{i+} + \textnormal{M} &\longrightarrow \textnormal{X}^{(i-q)+} + [\textnormal{M}]^{q+}, \label{eq:capture}
\end{align}
with "[M]" referring to the possibility for compound M to undergo, in the process, dissociation, excitation, and ionization, or a combination of these processes. Such a reaction for the solar wind is referred to as "solar wind charge exchange", abbreviated SWCX, to distinguish it from other charge-exchange processes involving, for instance, fast cometary ions and the slow-moving neutral gas. In the latter case, in the inner coma of a comet, where the solar wind is expected to be significantly slowed down as is the case for high-activity comets (such as 1P/Halley), charge exchange between cometary ions and neutrals may also play a role in converting fast cometary ions, which have been accelerated by local electric fields, into slow ones, potentially ensuring that the cometary plasma moves with the neutrals. 

Because the neutral atmosphere of a comet is in expansion and may extend to millions of kilometers in space \citep{Combi2004}, SWCX reactions may strongly affect the solar wind mass-loading upstream of the nucleus, and thus the formation and dynamics of bow shock and cometopause structures \citep{Gombosi1987,CSW2017}, and around the diamagnetic cavity \citep{Ip1990}. 
Moreover, the projectile solar wind ion is usually much faster ($\sim400$\,\kms{}) than the neutral species it hits ($\sim1$\,\kms{}), with the result that the charge-exchanged species mostly keep their initial kinetic energy \citep{CSW2018a}. Momentum transfer due to charge exchange between ion projectiles and neutral targets may also take place, and dominates at low impact energies the total momentum transfer cross sections \citep{Banks1973a,Ip1990}.

\begin{figure}
  \includegraphics[width=\linewidth]{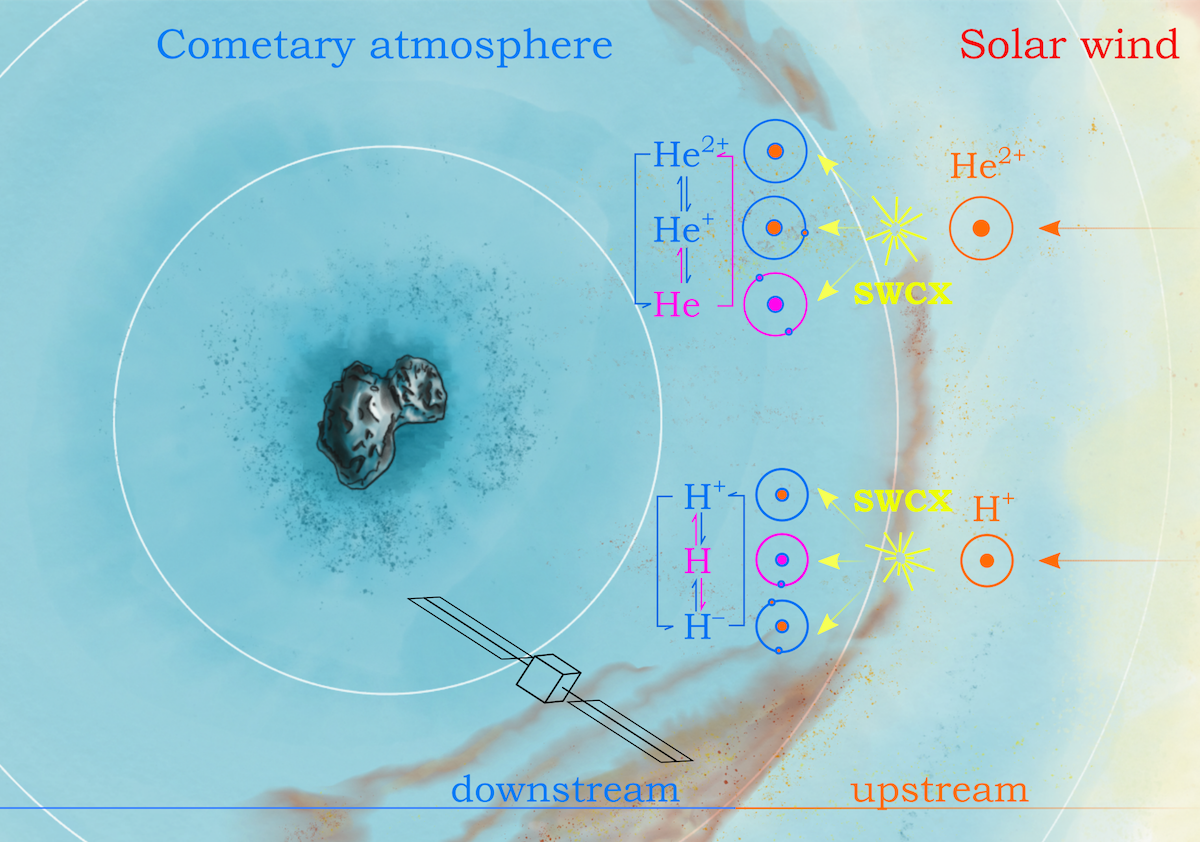}
     \caption{Sketch of solar wind charge-exchange interactions (SWCX) at comet 67P for a heliocentric distance of about $2$\,AU. The upstream solar wind, composed of H$^{+}$ and He$^{2+}$ ions, experiences charge-changing collisions when impacting the comet's neutral atmosphere, producing a mixture of charged states downstream of the collision. ENAs are depicted in pink. An increasingly deep blue color denotes a correspondingly denser atmosphere. SWCX plays a major role at and around the bow shock-like structure, depicted in shades of red, where the solar wind is heated and deflected.
     }
     \label{fig:cometSunsketch}
 \end{figure}

As solar wind H$^+$ and He$^{2+}$ ions impinge upon an atmosphere, charge-changing reactions start fractionating the initial charge state distribution into a mixture of H$^{+}$, H$^0$ and H$^-$ on the one hand, and of He$^{2+}$, He$^+$ and He$^0$ species on the other \citep{CSW2018b}. This effectively results in the production of fast energetic neutral atoms (ENAs), which are not bound to the plasma, but may interact further with the atmosphere downstream of the SWCX collision (see Fig.~\ref{fig:cometSunsketch}).

A spacecraft such as \emph{Rosetta,} which carried neutral, ion, and electron spectrometers deep in the atmosphere of comets or planets, can measure particle fluxes of individual charge states in situ \citep[see][]{Nilsson2007,burch2007ssr}. One such instrument is, for example, the Rosetta Plasma Consortium Ion Composition Analyser (RPC-ICA) on board \emph{Rosetta}. 
RPC-ICA is a top-hat mass spectrometer capable of simultaneously measuring the angular distribution of H$^+$, He$^{2+}$ , and He$^+$ ion fluxes with good temporal resolution \citep{Nilsson2015,Nilsson2015AA,Behar2016a,Behar2016b,Behar2017}. A wide range of studies has been performed with it and includes: evidence of SWCX and a simple analytical model \citep{CSW2016}, mass-loading of the solar wind \citep{Behar2016b,Behar2016a}, characterization of the solar wind ion cavity \citep{Behar2017}, high temporal resolution of ion dynamics \citep{StenbergWieser2017}, cometary ion dynamics \citep{Bercic2018}, characterization of tail plasma \citep{Behar2018tail}, and evidence for cometary bow shock detections \citep{Gunell2018,Alho2019}.

Another plasma instrument on board \emph{Rosetta} capable of studying SWCX reactions is the RPC Ion and Electron Sensor \citep[RPC-IES, ][]{burch2007ssr}. \cite{Burch2015} detected H$^-$ negative ions in the early phase of the mission for the first time. They were confirmed to be arising from two consecutive  single-electron captures from the initial solar wind protons \citep{Burch2015,CSW2018b}.

Complementing the pure plasma measurements, the neutral environment of the comet was probed using the mass spectrometers as part of the Rosetta Orbiter Spectrometer for Ion and Neutral Analysis \citep[ROSINA,][]{Balsiger2007}. This instrument package measured the precise neutral composition of the cometary atmosphere, linked it to precise outgassing regions on the nucleus \citep{Fougere2016,Lauter2018}, and painted a picture of the complex interplay between the main species at the comet, namely H$_2$O, CO$_2$, CO, and O$_2$ \citep{Bieler2015,Bieler2015nature}. Its ion channel was also able to distinguish among heavy ions (e.g., H$_2$O$^+$, H$_3$O$^+$, CO$_2^+$, and NH$_4^+$) in the ionosphere \citep{Fuselier2015,Fuselier2016,Beth2016}.

 The present work is the third of a series of studies on charge-changing reactions in cometary atmospheres, with application to the \emph{Rosetta} datasets.  
 It consists of three parts that we briefly describe below. 

\cite{CSW2018a} (hereafter Paper~I) presented a review of currently available experimental charge-changing and ionization cross sections of hydrogen and helium species in a water gas, with recommended low-energy values for typical solar wind energies. As H$_2$O is the most abundant cometary neutral species during most of the \emph{Rosetta} mission \citep{Lauter2018}, only this species was considered. Laboratory data needs were identified, which are required to bridge the gaps in the current experimental results. Polynomial fits for the systems $(\textnormal{H}^+, \textnormal{H}, \textnormal{and H}^-)-\textnormal{H}_2\textnormal{O}$ and $(\textnormal{He}^{2+}, \textnormal{He}^{+},  \textnormal{and He})-\textnormal{H}_2\textnormal{O}$ were proposed. This cross section review is therefore not confined to comet-solar wind plasma interactions, but may hold interest for the astrophysics, biophysics, and atomic/molecular physics communities. 
    
\cite{CSW2018b} (hereafter Paper~II) developed a general analytical solution of the three-species system of helium and hydrogen, with implications specific to comets. The forward model expressions were given, and two inversions, one for deriving the outgassing rate of the comet, one for estimating the upstream solar wind flux from in situ ion observations, were proposed. Using the recommended cross sections of Paper~I, the dependence on heliocentric distance, cometocentric distance, and solar wind speed was explored. From geometrical considerations alone, predictions for the charge state distribution at comet 67P at the location of \emph{Rosetta} were given.
    
In the present study, we apply the analytical forward and inverse models of Paper~II to \emph{Rosetta}'s ion and neutral observations. From local RPC-ICA ion flux measurements, we then retrieve the total neutral outgassing rate of comet 67P, which we compare to ROSINA neutral measurements. We also attempt to derive an estimate of the upstream solar wind flux from local ion measurements, with a comparison to solar wind flux estimates propagated from Mars and Earth. Finally, we present a mission overview of local ion productions due to SWCX and compare them to photoionization and electron impact ionization. In the closing paragraphs, we give mission-wide predictions on the ENA environment at the comet.

\section{Model description}\label{sec:model}
In order to describe how charge-transfer and stripping reactions impacted the solar wind plasma when it encountered the neutral environment of 67P during the \emph{Rosetta} mission, we developed in Paper~II a general 1D analytical solution of the three-charge component fluid system of hydrogen and helium solar wind particles impacting a neutral atmosphere. We briefly summarize the salient points of interest for this study  here and refer to Paper~I \citep{CSW2018a} and Paper~II \citep{CSW2018b} for details.

The model is based on the fluid continuity equation for solar wind ions. Several simplifying assumptions were made: (i) stationarity, (ii) all particles are moving along the solar wind velocity, and (iii) all charge states of a solar wind particle have the same speed and path.

Inputs of the forward analytical model include charge-changing cross sections and cometary neutral density. Velocity-dependent charge-changing (and ionization) cross sections are tabulated in Paper~I for hydrogen and helium projectiles impinging on an H$_2$O gas target ($\text{six}$ charge-changing reactions are taken into account per projectile species). For simplicity, only H$_2$O neutral targets are considered in the model, although, as emphasized in the preceding section, CO$_2$ became the more abundant species after March 2016. Because CO$_2$ and H$_2$O have commensurate electron capture cross sections \citep[of about a factor $2$ , see][and Paper~II]{Bodewits2006}, choosing one neutral target over the other or a mix of them does not result in drastically different results, especially when the total outgassing rate is determined.

With these assumptions, the calculation of the charge state distribution at any cometocentric distance $r$ is found to only depend on the column density of the atmosphere, noted $\eta$ \citep{Beth2016}:
\begin{equation}
     \begin{split}
                \eta(r,\chi) &= \int_{r\cos\chi}^{+\infty} n_n(s)\, \drv s = \frac{Q_0}{4\pi \varv_0\ r} \frac{\chi}{\sin\chi} = n_n(r)\,r\,\frac{\chi}{\sin\chi}, \label{eq:columnDensity}
        \end{split}
\end{equation}
where $n_n(r) = Q_0/\left(4\pi\,\varv_0\,r^2\right)$ is the local neutral density in the hypothesis of a collisionless spherically symmetric neutral outgassing \citep[][when neglecting photodestruction processes]{Haser1957}, $Q_0$ is the cometary neutral outgassing rate, $\varv_0$ is the speed of the neutrals, and $\chi$ is the solar zenith angle. The column density can be expressed as $\eta(r,\chi) = \frac{Q_0}{\varv_0}\ \epsilon(r,\chi)$, with $\epsilon = \chi/\left(4\pi\,r\sin\chi\right)$ a function of the observation geometry only. Quantities $Q_0$ and $\varv_0$ can be derived from local observations made by instruments on board \emph{Rosetta}, such as ROSINA-COPS \citep{Hansen2016,Fougere2016}.

Because of its simple analytical form, inversions of the three-component six-reaction model are also possible. Two such formulas are given in Paper~II to determine (i) the total neutral outgassing rate from local measurements of the $\textnormal{He}^+/\textnormal{He}^{2+}$ ion flux ratio $\mathcal{R}$, and (ii) the solar wind upstream density from ion measurements. Their form is recalled below for convenience. For outgassing rate $Q_0$, using particle flux ratio $\mathcal{R}(r_\textnormal{pos})=F(\textnormal{He}^+)/F(\textnormal{He}^{2+})=F_1/F_2$ measured at position $r_\textnormal{pos}$, we have
\begin{align}
     Q_0 &= \frac{\varv_0}{\epsilon(r,\chi)} \frac{\ln\left(\frac{N_1-\mathcal{R} N_2}{P_1-\mathcal{R} P_2}\right)}{2q},                  \label{eq:outgassingRetrieved}
\end{align}
where $(P_1,N_1)$, $(P_2,N_2)$, and $q$ are all functions of the six charge-changing cross sections $\sigma_{ij}(U_i)$, with $i$ and $j$ the initial and final charge states and $U_i$ the mean ion speed of species $i$. In the case of helium particles, subscripts $1$ and $2$ refer to charge states He$^+$ and He$^{2+}$, respectively. This formula is marginally similar to that derived in \cite{CSW2016}, when only one electron capture process (He$^{2+}\rightarrow$He$^+$, cross section $\sigma_{21}$) was taken into account:
\begin{align}
                Q_0 &= \frac{\varv_0}{\epsilon(r,\chi)} \frac{\ln\left(1 + \mathcal{R}\right)}{\sigma_{21}}.\label{eq:outgassingCSW2016}
\end{align}
This is itself a particular case of the electron loss-free solution presented in Appendix~B of Paper~II:
\begin{align}
    Q_0 & = \frac{\varv_0}{\epsilon(r,\chi)}\  \frac{\ln\left(1+ \frac{\sigma_{21}+\sigma_{20}-\sigma_{10}}{\sigma_{21}}\,\mathcal{R}\right)}{\sigma_{21}+\sigma_{20}-\sigma_{10}}.
    \label{eq:outgassingNoEL}
\end{align}

Differences between outgassing rates calculated from Eqs.\,(\ref{eq:outgassingRetrieved}), (\ref{eq:outgassingCSW2016}), and (\ref{eq:outgassingNoEL}) are discussed in Sect.\,\ref{sec:results}.

For the solar wind upstream particle flux retrieval $F_i^\textnormal{sw}$, we recall the general solution for the full system:
\begin{align}
    \quad &F_i^\textnormal{sw} = \frac{F_i(r_\textnormal{pos})}{F_i^{\infty} + \frac{1}{2q}\left( P_i\ e^{q \eta} - N_i\ e^{-q \eta} \right)\ e^{-\frac{1}{2} \sum{\sigma_{ij}(U_i)}\,\eta}},\label{eq:SWRetrieved}
\end{align}
where $F_i^\infty$ is the equilibrium flux or charge state $i$ and $F_i(r_\textnormal{pos})$ the ion flux measured locally. Subscript $i$ may refer here to H$^+$ (charge state $1$) or He$^{2+}$ (charge state $2$). The explicit parameterizations are given in Sects.\,2.2 and 2.3 of Paper~II.

In agreement with \cite{CSW2016}, the locally measured He$^+$/He$^{2+}$ particle flux ratio is a good proxy of the SWCX efficiency in a cometary coma.
The two inversions using RPC-ICA ion instrument flux measurements are presented in Sect.\,\ref{sec:results}, which are then compared to total outgassing rates from ROSINA-COPS and to interplanetary solar wind measurements from the satellites ACE and Mars Express.

In Paper~II, the sensitivity study to cometary parameters showed that single-electron capture of protons and $\alpha$ particles was the dominant process for typical solar wind speeds ($400-800$\,\kms{}), with double-electron capture of He$^{2+}$ playing the major role below $200$\,\kms{} however. Moreover, solar wind Maxwellian temperature effects start to play a role for temperatures above about $3\times10^6$\,K. For high outgassing rates at a constant $400$\,\kms{} solar wind speed, this would result in substantially enhanced fluxes of H$^+$  (SWCX is less efficient), whereas the opposite trend holds for He$^{2+}$ ions (SWCX becomes more efficient).

The model validity, depending on cometocentric and heliocentric distance, was discussed in Paper~II. It was estimated to range from a few tens of kilometers from the nucleus to large cometocentric distances, and depends on the actual cometary and solar parameters. Parameters such as outgassing rate, neutral density distributions, solar extreme-ultraviolet (EUV) flux, and solar wind parameters, as well as the position of \emph{Rosetta} during its orbit phase around comet 67P, varied greatly throughout the mission. In particular, the spatial asymmetry of the neutral environment and that of the plasma may both combine to alter where the model is expected to be valid or not. For case studies with quantitative comparison on the temporal scales of the cometary rotation period, these intricacies would need to be carefully examined. However, in the interpretation of charge-changing and ionization processes during the \emph{Rosetta} mission, as presented in this study, we show that the model may give access to reasonable estimates of the neutral outgassing rates and of the ENA environment.

\section{Rosetta solar wind measurements}\label{sec:RosettaData}
We present here the \emph{Rosetta} ion and total neutral density measurement, which we use in the forward and inverse analytical models.

\subsection{ Rosetta Ion Composition Analyser}\label{sec:RPC-ICA}
The RPC Ion Composition Analyser (RPC-ICA) was a top-hat ion mass spectrometer on board \emph{Rosetta} designed to image the 3D velocity distribution function of positive ions (solar wind and cometary) from $10$\,eV/charge to $40$\,keV/charge in $96$ channels ($\Delta E/E = 0.07$) \citep{Nilsson2007}. For a description of the instrument capabilities and results during the \emph{Rosetta} mission, we refer to \cite{Nilsson2015,Nilsson2015AA} (pre-perihelion solar wind and cometary ion mission overview) and \cite{Nilsson2017a} (full mission overview of the cometary plasma environment).  
The instrument had a $90^\circ\times360^\circ$ field of view and performed a complete angular scan in $192$\,s, which thus corresponds to the maximum temporal resolution of the instrument in the full energy mode. 

During the \emph{Rosetta} mission and approaching perihelion, the solar wind experienced an increased angular deflection with respect to the Sun-comet line, defining the formation of a so-called solar wind ion cavity, or SWIC \citep[][]{Behar2017}. This cavity, mostly free of solar wind ions, is thought to arise because of greater cometary outgassing activity and mass loading; in the \emph{Rosetta} RPC-ICA and RPC-IES datasets, it lasted from May to December 2015, in which almost no ions of solar wind origin could be detected. The cavity position with respect to the spacecraft is expected to be highly dynamic because of the solar wind variability and dynamics.
Thus, throughout the \emph{Rosetta} mission and outside of the SWIC, RPC-ICA routinely detected three main solar wind ions: H$^+$, He$^{2+}$, and charge-exchanged He$^+$ ions.

\textbf{Solar wind characteristics}. Solar wind velocity distribution moments are described in \cite{Behar2017}.
The ion density $n_\mathrm{sw}$ is the moment of order 0, and the ion bulk velocity $\vec{u}_\mathrm{sw}$ (a vector) appears in the moment of order 1, the flux density $n_\mathrm{sw} \ \underline{\mathbf{u}_\mathrm{sw}}$. The bulk speed can be defined as the norm of the bulk velocity, that is, $u_\mathrm{sw} = |\underline{\vec{u}_\mathrm{sw}}$|. However, this bulk speed is representative of single-particle speeds as long as the velocity distribution function is compact (e.g., a Maxwellian distribution). Complex velocity distribution functions were observed by RPC-ICA within the atmosphere of 67P. For instance, partial ring distributions were frequently observed for solar wind protons at intermediate heliocentric distances, when the spacecraft approached the SWIC \citep{Behar2017}. To illustrate the effect of such distorted distributions, a perfect ring (or shell) distribution centered on the origin of the plasma reference frame can be imagined, in which all particles have the same speed of $400$\,\kms{}. The norm of the bulk velocity in this case would be $0$\,\kms{}, whereas the mean speed of the particles is $400$\,\kms{}, which is the relevant speed for SWCX processes. This mean speed, noted $U_\mathrm{sw}$, of the particles is calculated by first summing the differential number flux over all angles, and then taking the statistical average \citep{Behar2018thesis}. Over the entire mission, the deceleration of the solar wind using the mean speed of the particles is much more limited than the deceleration shown by the norm of the bulk velocity \citep{Behar2017}: there is more kinetic energy in the solar wind than the bulk velocity vector would let us think. This is the main difference with the paradigm used at previously studied (and more active) comets \citep{Behar2018aa_model}. These complex, nonthermal velocity distribution functions also prevent us from reducing the second-order moment (the stress tensor\emph{}) to a single scalar\emph{} value, which, for a Maxwellian distribution, could be identified with a plasma temperature. In the context of 67P and for an important part of the cometary orbit around the Sun, the temperature of the solar wind proton has no formal definition.

Solar wind ion fluxes were for the first time calculated self-consistently as $F_\textnormal{sw} = n_\textnormal{sw}\,U_\textnormal{sw}$ , where $U_\textnormal{sw}$ is the mean solar wind speed; they are in excellent agreement with the integrated ion fluxes directly measured by RPC-ICA ($\sim10-15\%$ difference on average).

\begin{figure*}
  \includegraphics[width=\linewidth]{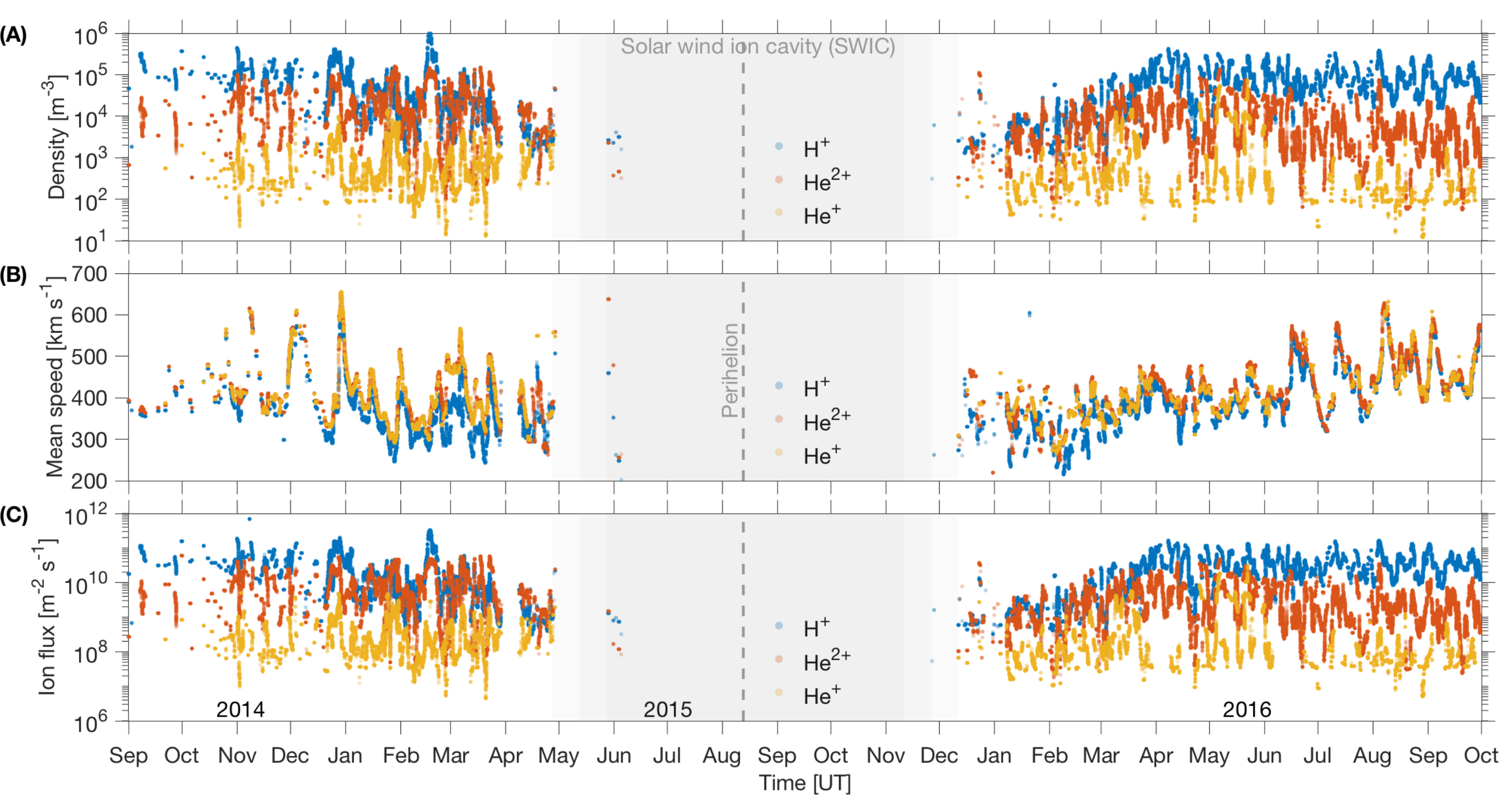}
     \caption{RPC-ICA-derived characteristics of the local solar wind ion distribution functions at comet 67P during the \emph{Rosetta} mission (2014-2016) for H$^+$, He$^{2+}$ , and He$^+$. (A) Density $n_i$, (B) mean ion speed $U_i$, and (C) ion flux $F_i$. The SWIC is approximately indicated as a gradually deeper gray region; its position with respect to the spacecraft is likely highly variable. All quantities have been submitted to a one-day moving average to remove variations due to the nucleus rotation. Increasingly intense colors indicate an increasing number of data points. 
     }
     \label{fig:ICAmoments}
 \end{figure*}

Figure~\ref{fig:ICAmoments} presents the RPC-ICA day-averaged densities (moment of order 0) (A), mean ion speeds $U_\mathrm{sw}$ (B), and ion fluxes $F_\mathrm{sw}$ (C) for H$^+$, He$^{2+}$, and He$^+$. Averages over $24$\,h are expected to remove variations due the rotation of the nucleus ($\sim12.4$\,h). Protons were usually the most abundant ions throughout the \emph{Rosetta} mission; they reached densities of $10^5$\,m$^{-3}$ on average until January 2015, whereas the helium species were one to two orders of magnitude less dense. After March 2015, a pronounced density decrease can be seen for H$^+$ and He$^{2+}$, which indicates that the spacecraft entered the SWIC. This is marked as a gray-shaded region in the figure.  In a mirror-like behavior, when the spacecraft left the SWIC in December 2015, densities started to increase again by about two orders of magnitude to their pre-SWIC levels, which are close to undisturbed solar wind levels at large heliocentric distances. This behavior was seen to a much lesser extent for He$^+$ ions, especially in the post-perihelion time frame, which is in part due to a much less favorable signal-to-noise ratio (and larger day-to-day fluctuations) for this somewhat scarce species. One question is thus whether charge-exchange reactions might be responsible for such a large decrease in particle flux and density when closing in on the SWIC. This question is further discussed in Sect.\,\ref{sec:upstreamSW}, when we retrieve the solar wind upstream conditions.

Mean ion speeds have all very similar values. The variations extend from $250$\,\kms{} to $650$\,\kms{} on average for all species, and proton speeds are marginally lower than those of helium particles.

Because the mean speeds of all ions are in agreement, a similar assessment can be made for the calculated ion fluxes as for the densities: a decrease of one to two magnitudes in fluxes when the spacecraft approached the SWIC, which was mirrored by a similar flux increase when \emph{Rosetta} exited it.

\subsection{ROSINA-COPS}
The Rosetta Orbiter Spectrometer for Ion and Neutral Analysis (ROSINA) is a suite of neutral and ion sensors \citep{Balsiger2007}. It consists of three instruments, one of which is the COmet Pressure Sensor (COPS), itself consisting of a nude and a ram gauge measuring the density and dynamic of the outflowing neutral gas, regardless of the composition \citep{Bieler2015}.  Another instrument of ROSINA is the Double Focusing Mass Spectrometer (DFMS): during the \emph{Rosetta} mission, it measured the abundance of cometary volatiles \citep{Fougere2016}. By analyzing DFMS and COPS data with a 3D Monte Carlo direct simulation model, \cite{Hansen2016} developed an empirical model of the neutral coma; they also performed fits on the water production rate $Q_0$ for inbound (pre-perihelion) and outbound (post-perihelion) portions of the orbit around the Sun, corrected for seasonal effects (due to the changing latitude/longitude coverage of the spacecraft). Their finding is recalled here.
\begin{equation}
Q_0 =
    \begin{cases}
            (2.58\pm0.12)\times 10^{28}\ R_\textnormal{Sun}^{-5.10\pm0.05}\ \textnormal{s}^{-1} \quad\text{inbound},  \\
            (1.58\pm0.09)\times 10^{29}\ R_\textnormal{Sun}^{-7.15\pm0.08}\ \textnormal{s}^{-1} \quad\text{outbound}.
    \end{cases}
    \label{eq:outgassingHansen}
\end{equation}

Recently, \cite{Lauter2018} studied the temporal evolution of the H$_2$O, CO$_2$, CO, and O$_2$ neutral densities and outgassing rates using COPS and DFMS. In accordance with \cite{Fougere2016}, they showed that H$_2$O dominated the neutral production in the coma during most of the mission, except after the post-perihelion equinox (March 2016, around $2.5$\,AU), when CO$_2$ became the main outgassed species.

In the simple assumption of a spherically expanding cometary atmosphere \citep[following][]{Haser1957}, the equivalent local outgassing rate inferred from ROSINA-COPS can be calculated as
\begin{align}
        Q_\textnormal{ROS} & = 4\pi\ \varv_0\ r^2\ n_\textnormal{ROS},\label{eq:outgassingROSINA}
\end{align}
with $n_\textnormal{ROS}$ the neutral gauge density in m$^{-3}$ measured by ROSINA-COPS at the location of the spacecraft, $\varv_0$ the speed of the neutral gas in m~s$^{-1}$, and $r$ \emph{Rosetta}'s cometocentric distance in m.
In the remaining study and for simplicity, we use the empirically determined analytical function of \cite{Hansen2016} to calculate the radial speed of the neutral gas $\varv_0$, in m\,s$^{-1}$:
\begin{align}
        \varv_0    &= \left(m_{R} R_\textnormal{Sun} + b_{R}\right)\ \left(1 + 0.171\ e^{-\frac{R_\textnormal{Sun}-1.24}{0.13}}\right),\label{eq:neutralVelocity}\\
    \textnormal{with}\quad m_{R} &= -55.5\quad\textnormal{and}\quad b_{R} =  771.0,\nonumber
\end{align}
where $m_R$ and $b_R$ are fitting parameters, and $R_\textnormal{Sun}$ is the heliocentric distance, expressed in AU. 

The total ROSINA-COPS outgassing rate $Q_\textnormal{ROS}$, regardless of species involved, is shown in Fig.~\ref{fig:outgassingrate} (black line). The inbound/outbound fits of \cite{Hansen2016} are shown as an orange line in the same plot.

\begin{figure*}
  \includegraphics[width=\linewidth]{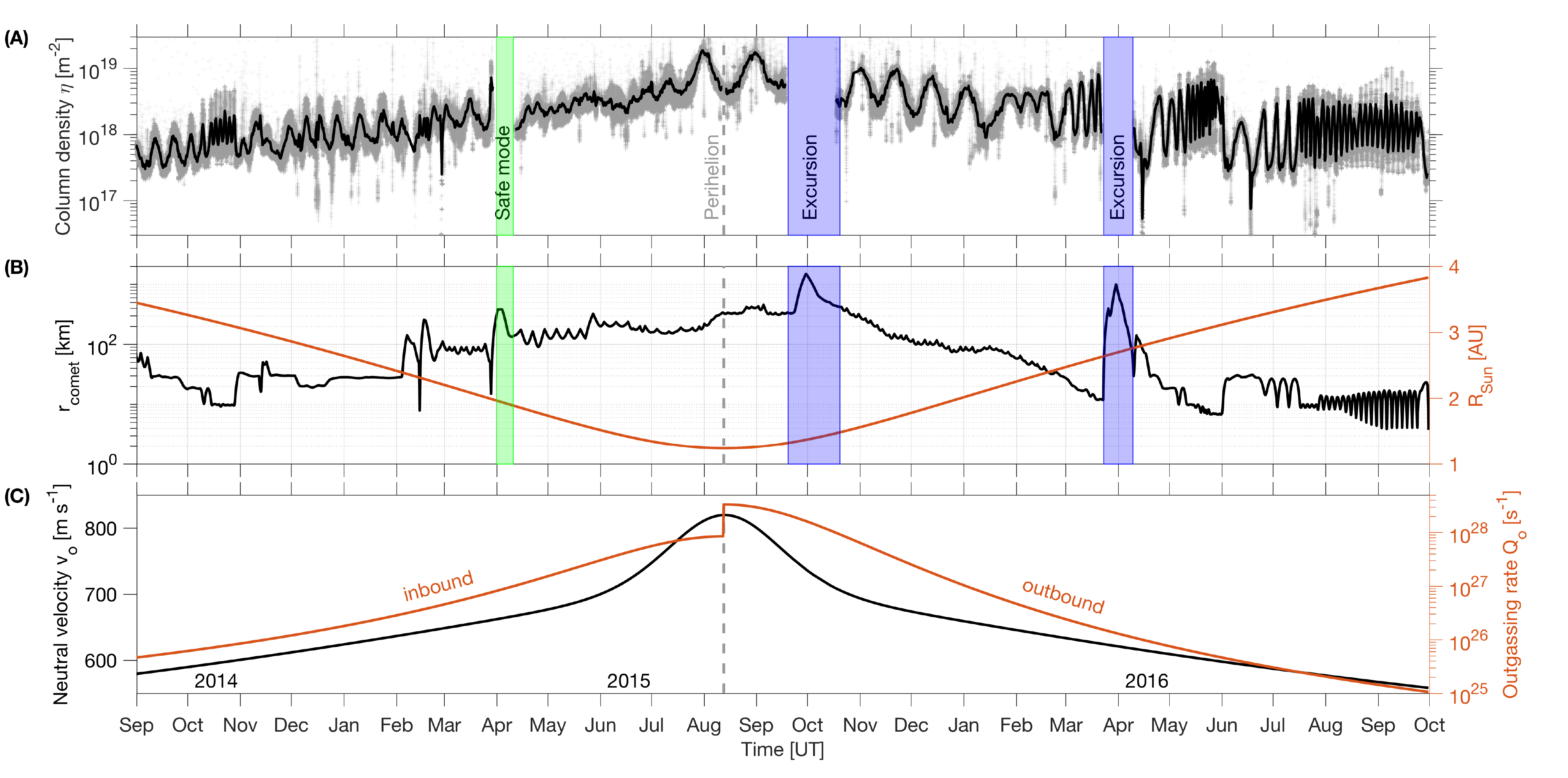}
     \caption{ROSINA-COPS-derived local neutral measurements at comet 67P during the \emph{Rosetta} mission. (A) Time series of the local estimated "upstream" column density of neutral species (gray crosses) from Eq.~(\ref{eq:columnDensity}), and 24h moving-averaged values (black line). (B) Cometocentric distance of \emph{Rosetta} (left), and heliocentric distance of comet 67P during the mission. (C) Empirically derived neutral outgassing speed (left axis) and outgassing rate (right axis) from Eqs.~(\ref{eq:neutralVelocity}) and (\ref{eq:outgassingHansen}). Safe mode and excursions are indicated.
     }
     \label{fig:ROSINA}
 \end{figure*}

The total neutral column density $\eta$, empirical gas speeds $\varv_0$ , and outgassing rate $Q_0$ from ROSINA-COPS are shown in Fig.~\ref{fig:ROSINA}A and C throughout the \emph{Rosetta} mission. To calculate $\eta$, we used Eq.~(\ref{eq:columnDensity}), as in \cite{Beth2016}.
Periods when the spacecraft was in "safe mode" ($1-10$ April 2016), as well as the two excursions at large cometocentric distance, one in late-September to mid-October 2015 (dayside excursion) and another in late-March  to early-April (cometary tail excursion), are shown (Fig.~\ref{fig:ROSINA}B). These periods correspond to times when ROSINA measurements yielded large uncertainties or when the sensor was turned off. These periods are ignored in our interpretation. The column density varies by two orders of magnitude between perihelion ($1.24$\,AU, $\eta\approx2\times10^{19}$\,m$^{-2}$) and large heliocentric distances ($3$\,AU, $\eta\approx3\times10^{17}$\,m$^{-2}$), depending on \emph{Rosetta}'s location at the comet throughout the mission. Oscillations in the column density, which for instance appear on a monthly basis after perihelion, are due to spacecraft latitudinal changes and different nucleus activities between the northern and southern hemispheres \citep[see][]{Heritier2018}. Predictably, neutral speeds vary between about $550$ and $850$\,m~s$^{-1}$, and outgassing rates vary between about $10^{25}$ and $3\times10^{28}$~s$^{-1}$, with maxima reached around perihelion.

\section{Results and discussion}\label{sec:results}
The helium and hydrogen charge state distributions can be reconstructed using the analytical model presented in Paper~II and based on the charge-changing and ionization cross section survey presented in Paper~I. 
Such a reconstruction is used to help interpret the mission-wide ion spectrometer datasets. It was made in two steps: (i) the solar wind mean ion speed measured at the comet by RPC-ICA was used to derive the relevant energy-dependent cross sections throughout the mission, 
(ii) using the analytical model in forward or in inverse mode, the neutral outgassing rate and the upstream solar wind conditions may be successively inferred, and serve as tests of the validity of the model. Other interesting points, such as a summary of all ionization processes at the comet and an overview of the ENA environment of comet 67P can be derived, and are finally presented.

\subsection{Local cross sections during the mission}
In this study, charge-changing cross sections are presented for helium and hydrogen during the \emph{Rosetta} mission. They are based on the careful survey made in Paper~I and on our recommended velocity-dependent polynomial fits. 
We calculated the cross sections using the mean ion speed measured by RPC-ICA (Sect.\,\ref{sec:RPC-ICA} and Fig.~\ref{fig:ICAmoments}B). For protons, we used $U_\textnormal{sw}(\textnormal{H}^+)$. For all helium ions, we used for simplicity $U_\textnormal{sw}(\textnormal{He}^{2+})$, because $U_\textnormal{sw}(\textnormal{He}^{2+})$ and $U_\textnormal{sw}(\textnormal{He}^{+})$ have similar values and variations throughout the mission. The obtained charge-changing cross-section uncertainties varied with the charged state considered: they were $<25\%$ for He$^{2+}$ and He$^+$, but were $>75\%$ for He. Cross sections for hydrogen particles have large uncertainties on average, usually $>50\%$. For a thorough discussion of uncertainties and the reliability of the polynomial fits, we refer to Paper~I.

\begin{figure*}
  \includegraphics[width=\linewidth]{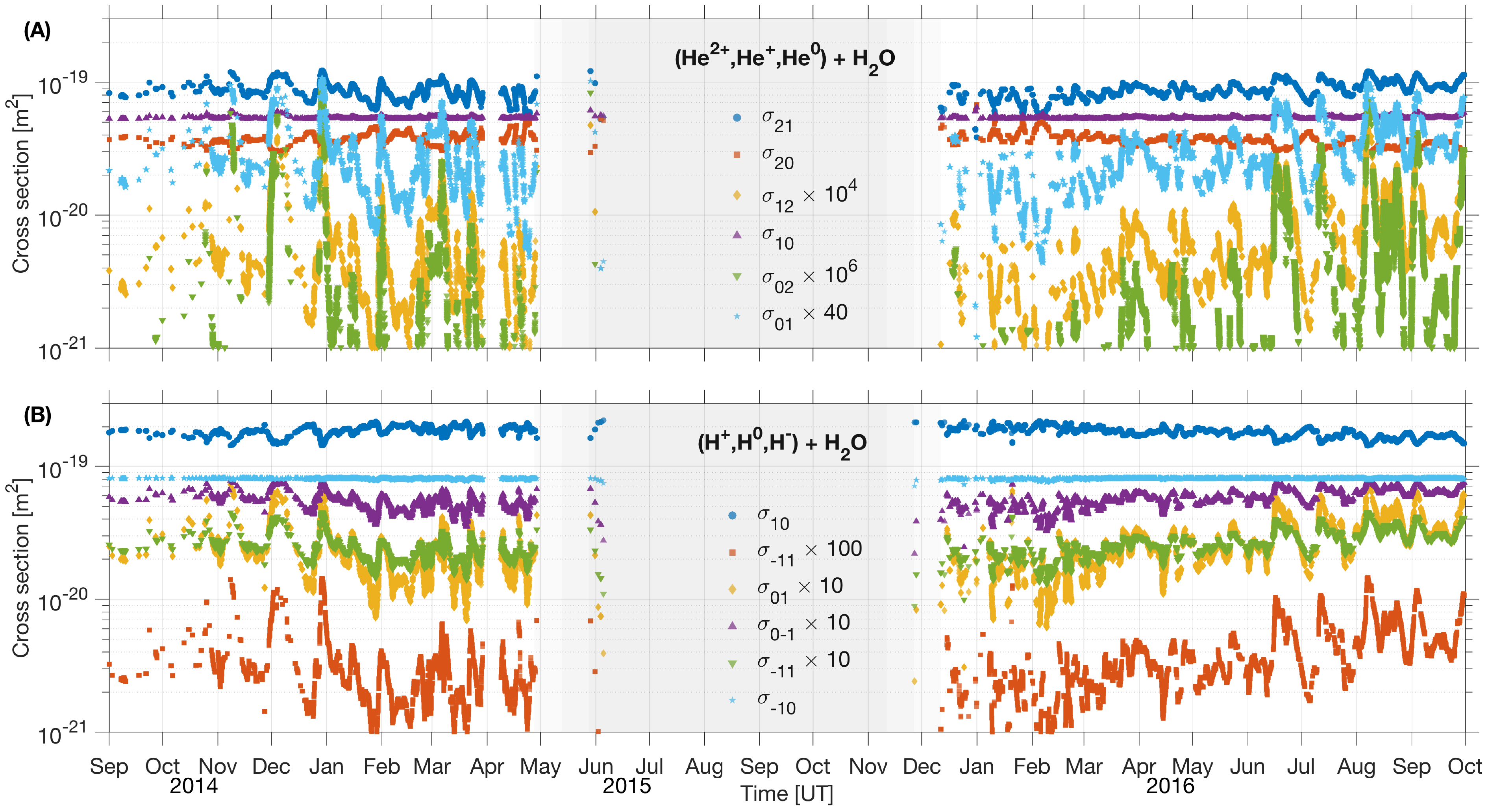}
     \caption{Monochromatic charge-changing cross sections during the \emph{Rosetta} mission (2014-2016). (A) Helium species. (B) Hydrogen species. 
     Cross sections are calculated using the mean solar wind ion speed of particles measured by RPC-ICA, with a solar wind temperature of $T=0$\,K. A one-day moving average was performed for clarity. The solar wind ion cavity is indicated as a gray-shaded region.
     }
     \label{fig:XsectionsMission}
 \end{figure*}

Figure~\ref{fig:XsectionsMission} shows the monochromatic charge-changing cross sections calculated at the mean speed of the solar wind, measured by RPC-ICA. Because no significant deceleration of the solar wind is seen in the RPC-ICA data throughout the mission, the thermal speed of the solar wind can be considered negligible with respect to the solar wind speed (equivalent to $T=0$\,K). 

Of all charge-changing processes, single-electron captures by He$^{2+}$ and H$^+$ have the largest cross section throughout the mission, with values $\gtrsim10^{-19}$\,m$^{2}$. For helium, two other processes have high cross sections ($\gtrsim 4\times10^{-20}$\,m$^{2}$): the single-electron capture by He$^{+}$ ($\sigma_{10}^\textnormal{He}$), and the double-electron capture of He$^{2+}$ ($\sigma_{20}^\textnormal{He}$), especially between January 2015 and March 2016, when the heliocentric distance is shorter than $\sim2.5$\,AU. This results in the efficient conversion of He$^{2+}$ ions into He ENAs. For hydrogen, the single-electron loss of H$^-$ also has a large cross section, with $\sigma_{-10}^\textnormal{H}\approx8\times10^{-20}$\,m$^{2}$, which is a factor of about $2$ lower than $\sigma_{10}^\textnormal{H}$. All other processes involving hydrogen particles have almost negligible cross sections in comparison; this preferentially creates H ENAs from solar wind protons, with these ENAs only undergoing a few losses to H$^+$ or to H$^-$.

During most of the mission, solar wind ion speeds did not vary much, with an average speed around $400$\,\kms{}, and they never dropped to below $250$\,\kms{}. This implies that processes other than electron capture did not contribute much to changes in the charge state distribution. Consequently, when a Maxwellian-like solar wind is taken into account (with $T=40\times10^6$\,K for maximum effect, see Paper~I and Paper~II for a theoretical discussion), only minor changes are expected in the cross-section magnitudes, with the exception of ionization cross sections. In the following, we therefore only consider "monochromatic" cross sections, and not Maxwellian-averaged cross sections, unless specified otherwise.   

\subsection{Mission-wide inversions of \emph{Rosetta} datasets}
In this section, we present two inversions of the RPC-ICA datasets using our solar wind charge-changing analytical model. One results in the retrieval of the total cometary outgassing rate, the other in the reconstruction of solar wind upstream densities and fluxes.

\subsubsection{Cometary outgassing rate}\label{sec:outgassingRetrieval}
The inversion of our analytical model to retrieve the neutral outgassing rate from RPC-ICA ion flux data follows three steps: (i) calculation of $\mathcal{R}=F$(He$^+$)/$F$(He$^{2+}$) ion flux ratio measured by RPC-ICA locally, (ii) calculation of the geometrical quantity $\epsilon(r,\chi)=\chi/\left(4\pi\,\sin\chi\right)$ from the spacecraft position, and (iii) final inversion using expression~(\ref{eq:outgassingRetrieved}) for $Q_0$. In this step, we assume that the atmosphere is solely composed of H$_2$O. Consequently, we adopt the velocity-dependent charge-changing cross sections in H$_2$O from Paper~I, with solar wind speeds for hydrogen particles taken as $U_i = U_{\textnormal{H}^{+}}$ and for helium particles as $U_i = U_{\textnormal{He}^{2+}}$.

\begin{figure*}
  \includegraphics[width=\linewidth]{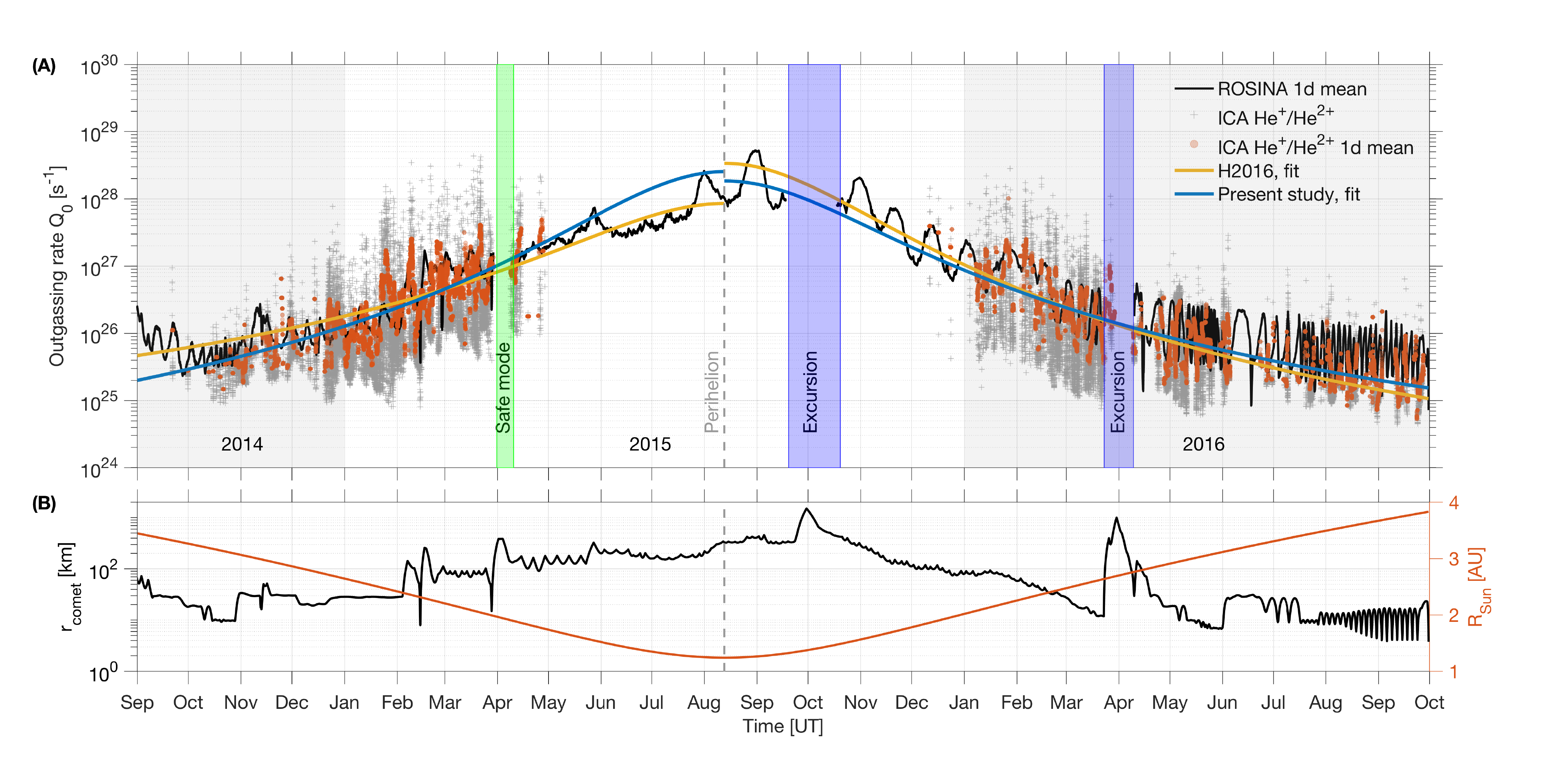}
     \caption{(A) Local water outgassing rate of comet 67P during the \emph{Rosetta} mission (2014-2016), as measured by ROSINA (black line, one-day moving average) 
     and retrieved from RPC-ICA. RPC-ICA one-day moving averages are presented as red circles, whereas the full non-averaged time series is shown as gray $\text{pluses}$. Safe mode and excursions are indicated: at these dates, the outgassing rate from ROSINA-COPS yields unreliable results. Inbound and outbound fits to the ROSINA data of \cite{Hansen2016} (orange line, corrected for latitude/longitude effects) and to the RPC-ICA data (blue line, this study) are shown. (B) Cometocentric (left axis) and heliocentric distances (right axis) during the mission. Instrumental uncertainties are estimated to be $15\%$ for ROSINA-COPS.
     }
     \label{fig:outgassingrate}
 \end{figure*}

Figure~\ref{fig:outgassingrate}A presents the local H$_2$O outgassing rate derived from in situ RPC-ICA measurements of ion flux ratios and compares it to those retrieved from ROSINA-COPS measurements using Eq.~(\ref{eq:outgassingROSINA}). 
We also superimpose the geometry-corrected fits of \cite{Hansen2016} for pre- and post-perihelion \emph{Rosetta} datasets. For reference, the cometocentric and heliocentric distance variations are included in Fig.~\ref{fig:outgassingrate}B. 

For the ROSINA-COPS and RPC-ICA measurements, we performed $24$h moving averages to suppress variations due to the rotation of the cometary nucleus.
Outgassing rates from ROSINA-COPS and from RPC-ICA are in very good agreement throughout the mission, and the neutral atmosphere variations are convincingly captured. On average, when a running average over 14 days is applied to study long-term trends, the RPC-ICA-derived rates are lower by a factor of about $1.3$ than those from ROSINA-COPS. 
Local differences between the two estimates may arise, for example, (i) from the ion dynamics and the increasing deflection and energy spread of solar wind ions, and (ii) from the assumption of a spherically symmetric outgassing used in this work. Such effects would need to be examined case by case and are therefore beyond the scope of this study. However, the improvement over past studies in the outgassing rate retrieval is clear; in comparison, using the simple inversion of \cite{CSW2016} and Eq.~(\ref{eq:outgassingCSW2016}) when one electron capture reaction is taken into account, the division factor stated above for 14-day running averages reaches about $1.8$ throughout the mission, in agreement with the results of \cite{CSW2016}. This indicates how relatively efficient other processes such as double-electron capture of He$^{2+}$ and single-electron capture of He$^{+}$ may become, depending on solar wind speed and cometocentric distance. Expression~(\ref{eq:outgassingNoEL}) for the outgassing rate, when no electron losses (no stripping reactions) are taken into account,  most of the time yields results in between those from Eqs.~(\ref{eq:outgassingRetrieved}) and (\ref{eq:outgassingCSW2016}): from January 2015 to June 2016, it remains within about $10\%$ of the full $\text{six}$-reaction model. Stripping reactions (especially $\sigma_{01}$ for He) are found to play an increasingly important role at large heliocentric distances (small cometocentric distances, at the validity limit of the model), where the no-electron-loss model produces outgassing rates that are on average $>30\%$ lower than the result of the full six-reaction model.

\begin{figure}
  \includegraphics[width=\linewidth]{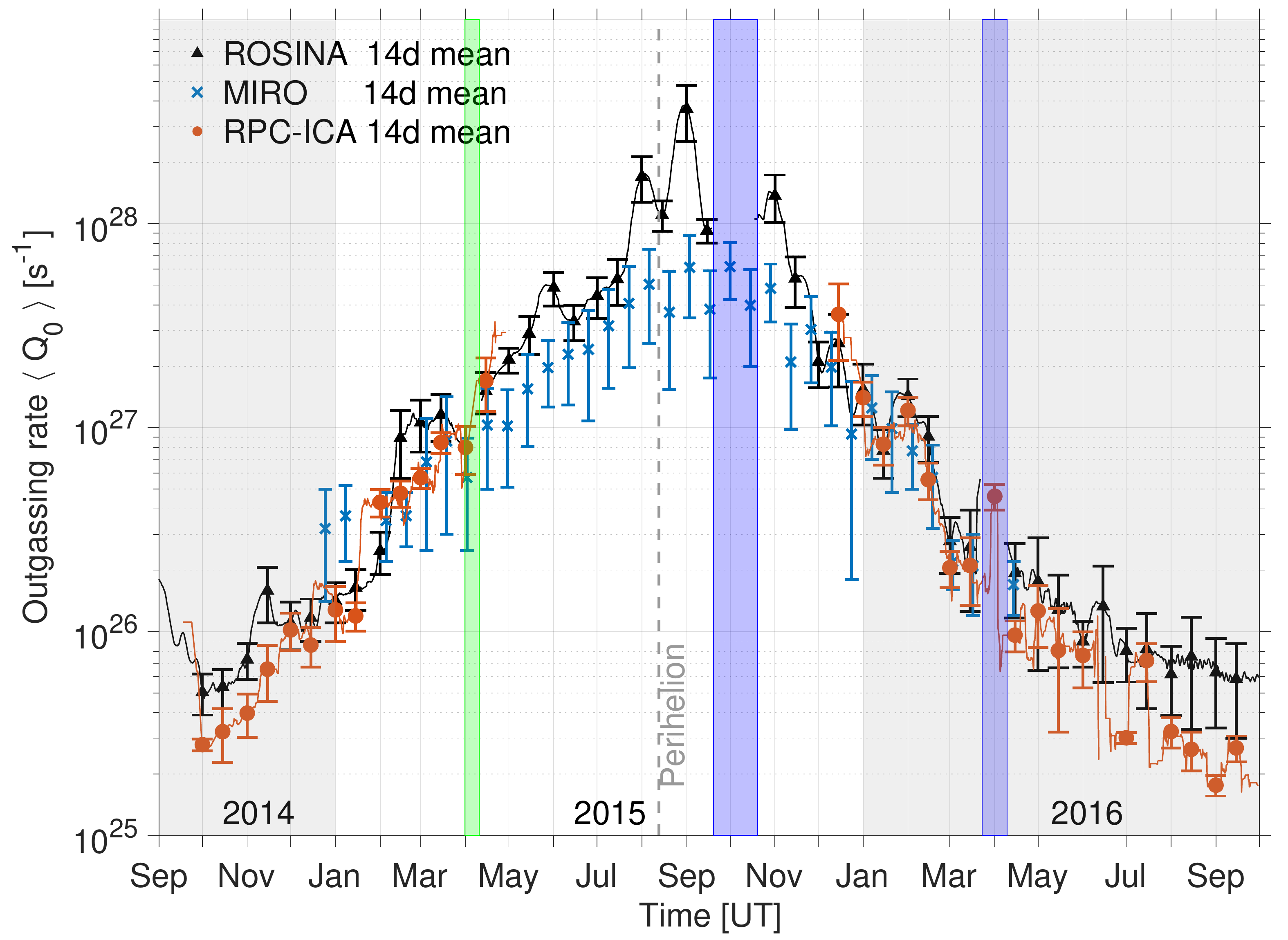}
     \caption{Local 14-day averaged water outgassing rate of comet 67P during the \emph{Rosetta} mission (2014-2016) as measured by ROSINA-COPS (black line and triangles), MIRO \citep[blue crosses, from][]{Marshall2017}, 
     and retrieved from RPC-ICA (red line and circles, tabulated in Table~\ref{tab:outgassingRates}). Error bars correspond to the median absolute standard deviations. Colored regions denote spacecraft excursions and safe modes, as in Fig.~\ref{fig:outgassingrate}.
     }
     \label{fig:outgassingrate14d}
 \end{figure}

To examine the long-term trends of our retrievals in detail, the 14-day moving average of the RPC-ICA-derived local H$_2$O outgassing rates, $\langle Q_0 \rangle$, is tabulated in Table~\ref{tab:outgassingRates} and plotted in Fig.~\ref{fig:outgassingrate14d}. Strong fluctuations of the measured  RPC-ICA signal result in large standard deviations that at times exceed $30\%$. We compare these rates to the $14$-day moving averages of the local H$_2$O outgassing rates of \cite{Marshall2017} (see their Table~1) using the Microwave Instrument for the Rosetta Orbiter (MIRO): these averages have errors larger than $30\%$ at large cometocentric distances and reach about $50\%$ at perihelion. The RPC-ICA rates compare well with those of MIRO in their common temporal coverage, as well as with those of ROSINA-COPS (black triangles in the figure). At large heliocentric distances, especially after July 2016 (post-perihelion), large differences between ROSINA-COPS and RPC-ICA can be seen: they coincide with \emph{Rosetta}'s increasingly deep dips into the inner cometary atmosphere around and below $10$\,km cometocentric distance. As recalled in Sect.\,\ref{sec:model}, this distance corresponds to the lower validity limit of the analytical model, and may at those times explain in part the poor quality of our outgassing rate retrieval.

  \begin{table}
        \centering
        \caption{Local water outgassing rates $\langle Q_0\rangle$ derived from RPC-ICA charge-exchange analysis of solar wind ion fluxes, with a $14$-day moving average applied to the full dataset, as shown in Fig.~\ref{fig:outgassingrate14d}. $R_\mathrm{Sun}$ is the heliocentric distance. 
        }
        \label{tab:outgassingRates}
    \tiny
        \begin{tabular}{lcc} 
                \hline\hline
        Date & Heliocentric distance  & Local average outgassing rate \\ \relax
                [UT] & $R_\mathrm{Sun}$ [AU] & $\langle Q_0\rangle$ [$\times10^{26}$\,s$^{-1}$] \\
                \hline
     & Pre-perihelion & \\
     01-10-2014 & 3.26 & $0.28\pm0.02$ \\
     15-10-2014 & 3.16 & $0.32\pm0.10$ \\
     01-11-2014 & 3.06 & $0.40\pm0.10$ \\
     15-11-2014 & 2.96 & $0.65\pm0.20$ \\
     01-12-2014 & 2.86 & $1.00\pm0.21$ \\
     15-12-2014 & 2.75 & $0.86\pm0.19$ \\
     01-01-2015 & 2.65 & $1.30\pm0.39$ \\
     15-01-2015 & 2.54 & $1.20\pm0.19$ \\
     01-02-2015 & 2.42 & $4.30\pm0.66$ \\
     15-02-2015 & 2.31 & $4.80\pm0.70$ \\
     01-03-2015 & 2.21 & $5.70\pm0.64$ \\
     15-03-2015 & 2.10 & $8.50\pm1.00$ \\
     01-04-2015 & 1.99 & $8.00\pm2.10$ \\
     15-04-2015 & 1.85 & $17.00\pm5.00$ \\
     & Post-perihelion & \\
     15-12-2015 & 1.91 & $36.00\pm15.00$ \\
     01-01-2016 & 2.01 & $14.00\pm2.70$ \\
     15-01-2016 & 2.12 & $8.30\pm1.70$ \\
     01-02-2016 & 2.25 & $12.00\pm1.90$ \\
     15-02-2016 & 2.37 & $5.60\pm1.10$ \\
     01-03-2016 & 2.48 & $2.10\pm0.42$ \\
     15-03-2016 & 2.58 & $2.10\pm0.77$ \\
     01-04-2016 & 2.70 & $4.60\pm0.67$ \\
     15-04-2016 & 2.80 & $0.96\pm0.16$ \\
     01-05-2016 & 2.91 & $1.30\pm0.43$ \\
     15-05-2016 & 3.00 & $0.81\pm0.49$ \\
     01-06-2016 & 3.12 & $0.76\pm0.23$ \\
     01-07-2016 & 3.30 & $0.30\pm0.02$ \\
     15-07-2016 & 3.39 & $0.72\pm0.15$ \\
     01-08-2016 & 3.50 & $0.32\pm0.06$ \\
     15-08-2016 & 3.58 & $0.26\pm0.06$ \\
     01-09-2016 & 3.67 & $0.18\pm0.02$ \\
     15-09-2016 & 3.75 & $0.27\pm0.04$ \\
                \hline
        \end{tabular}
        \tablefoot{Median absolute standard deviations of the 14-day variations are given as an indication of the variability of the particle flux ratio measured by RPC-ICA.}
 \end{table}

Other sources of errors in our retrievals are present. The charge-changing cross-section uncertainties are usually larger than $25\%$ (see Paper~I);  propagating these uncertainties in the model would result in different outgassing rate retrievals. Because of the number of involved cross sections and associated errors, and because of fluctuations in the RPC-ICA moment-derived fluxes in the first place, outgassing rate retrievals may change by an estimated $50\%$. This is arguably much less than the daily fluctuations of the signal, which can reach $>100\%$ because of varying solar wind and plasma conditions encountered at the comet.    

We described above that although CO$_2$ becomes the main neutral cometary species after March 2016, the retrieval of the total outgassing rate from RPC-ICA is in rather good agreement with that of ROSINA-COPS until the end of the mission, with an underestimation of the RPC-ICA outgassing rate after July 2016. Because charge-changing cross sections in H$_2$O and CO$_2$ only differ by a factor $2$ at most, this underestimate can thus be ascribed to at least two supplementary factors: (i) the changing composition of the atmosphere, and (ii) the fast-varying column densities encountered by \emph{Rosetta} at that time (see Fig.~\ref{fig:ROSINA}A).

Least-square fits of the form $\alpha\,R_\textnormal{Sun}^{-\beta}$ were performed on the one-day averaged RPC-ICA data, with the following results:
\begin{equation}
Q_{\textnormal{ICA}} =
    \begin{cases}
            (1.16\pm0.04)\times 10^{29}\ R_\textnormal{Sun}^{-7.00\pm0.03}\ \textnormal{s}^{-1} \quad\text{inbound},  \\
            (7.21\pm0.14)\times 10^{28}\ R_\textnormal{Sun}^{-6.29\pm0.02}\ \textnormal{s}^{-1} \quad\text{outbound}.
    \end{cases}
    \label{eq:outgassingICA}
\end{equation}

The errors correspond to the mean one-day variations of the signal and therefore do not include the instrument uncertainty. These fits are shown in Fig.~\ref{fig:outgassingrate} (blue lines). At perihelion, our inbound and outbound fits connect rather well, with only a slight discontinuity. Although on average these fits do bear similarities with the season-corrected fits of \cite{Hansen2016}, they do differ, predominantly outside of times when observations from RPC-ICA and ROSINA-COPS were simultaneously available, which is expected. Except for the temporal coverage, one main difference is that \cite{Hansen2016} first corrected the ROSINA-COPS data for seasonal effects by applying their DSMC model before they performed the fits. Our approach is much simplified in comparison, as only $24$-hour means (almost two nucleus rotations) were made on the RPC-ICA data before the fits were performing. For the later part of the mission (after the second large-distance excursion), our fits are arguably in better agreement with ROSINA-COPS data than the empirical fits of \cite{Hansen2016}: the reason is that the post-excursion observations were not yet available at the time of publication of the \cite{Hansen2016} study. A reverse conclusion holds for example in the early stages of the mission (before October 2014), when RPC-ICA could not detect He$^{+}$ ions because of poor temporal coverage due to instrument problems \citep{Nilsson2015,Nilsson2017a}. In that period, our fit therefore substantially underestimates the measured ROSINA-COPS outgassing, which is better captured by the model of \cite{Hansen2016}.

We note that our inbound fit parameterization is coincidentally very close to the value determined by \cite{CSW2016} ($1.14\times10^{29}\,R_\textnormal{Sun}^{-7.06}$), where events were manually selected and particle fluxes were energy-summed individually (instead of calculating the moments of the ion distribution functions), and where only one charge-exchange reaction was used to invert the particle flux ratios. When this one-reaction simplification is applied to our current moment-based dataset, the fit becomes $Q_0 = 1.39\times10^{29}\,R_\textnormal{Sun}^{-7.58}$ for the inbound leg, which is a factor $1.5$ to $2$ lower for $R_\textnormal{Sun}>2.5$\,AU (when \emph{Rosetta} orbited at cometocentric distances smaller $50$\,km) than our current estimate with the full analytical six-reaction model. A similar conclusion befits the outbound leg data, for which the simple one-reaction model yields $Q_0 = 1.58\times10^{29}\,R_\textnormal{Sun}^{-7.44}$ (factor $1.4$ to $2.2$ lower for $R_\textnormal{Sun}>2.5$\,AU), which is caused by the progressively closer probing of \emph{Rosetta} to the nucleus after May 2016. In agreement with Paper~II, this result indicates that other processes than single-electron capture by He$^{2+}$ became more and more important at orbits that lay deeper inside the coma. 

Obtaining a lower limit estimate of the outgassing rate is, in principle, possible by studying the appropriate ion cyclotron wave (ICW) activity at the comet with the use of a magnetometer, as demonstrated by \citet{Huddleston1998} and \citet{Volwerk2001, Volwerk2013a,Volwerk2013b}. Such a magnetometer is present on board \emph{Rosetta} as part of the Rosetta Plasma Consortium, the RPC-MAG fluxgate magnetometer \citep{glassmeier2007ssr}. Because of the low-frequency nature of these waves, $5\times10^{-3}~\leq~f_{\rm c,H2O}~\sim~2\times10^{-2}$\,Hz and the strong compressional wave power, detection is difficult. A detailed study is currently under way.

\subsubsection{Upstream solar wind}\label{sec:upstreamSW}
 As discussed in Paper~II, SWCX is responsible for the conversion of upstream solar wind He$^{2+}$ ions into a mixture of He$^{2+}$, He$^{+}$ , and He$^{0}$ particles: this is expected to result in the local decrease of solar wind He$^{2+}$ ion fluxes measured deep in the cometary coma. The question now is to quantify this decrease. In Sect.\,\ref{sec:model} we recalled that it is possible to retrieve  the upstream solar wind fluxes from local RPC-ICA measurements of H$^+$ and He$^{2+}$ fluxes. For this purpose, we inverted the analytical model and used Eq.~(\ref{eq:SWRetrieved}). To be consistent in our approach, we used as inputs the locally measured He$^{2+}$ fluxes, the RPC-ICA-derived outgassing rates retrieved in the previous section, and the neutral velocity reported by \cite{Hansen2016}.

\begin{figure*}
  \includegraphics[width=\linewidth]{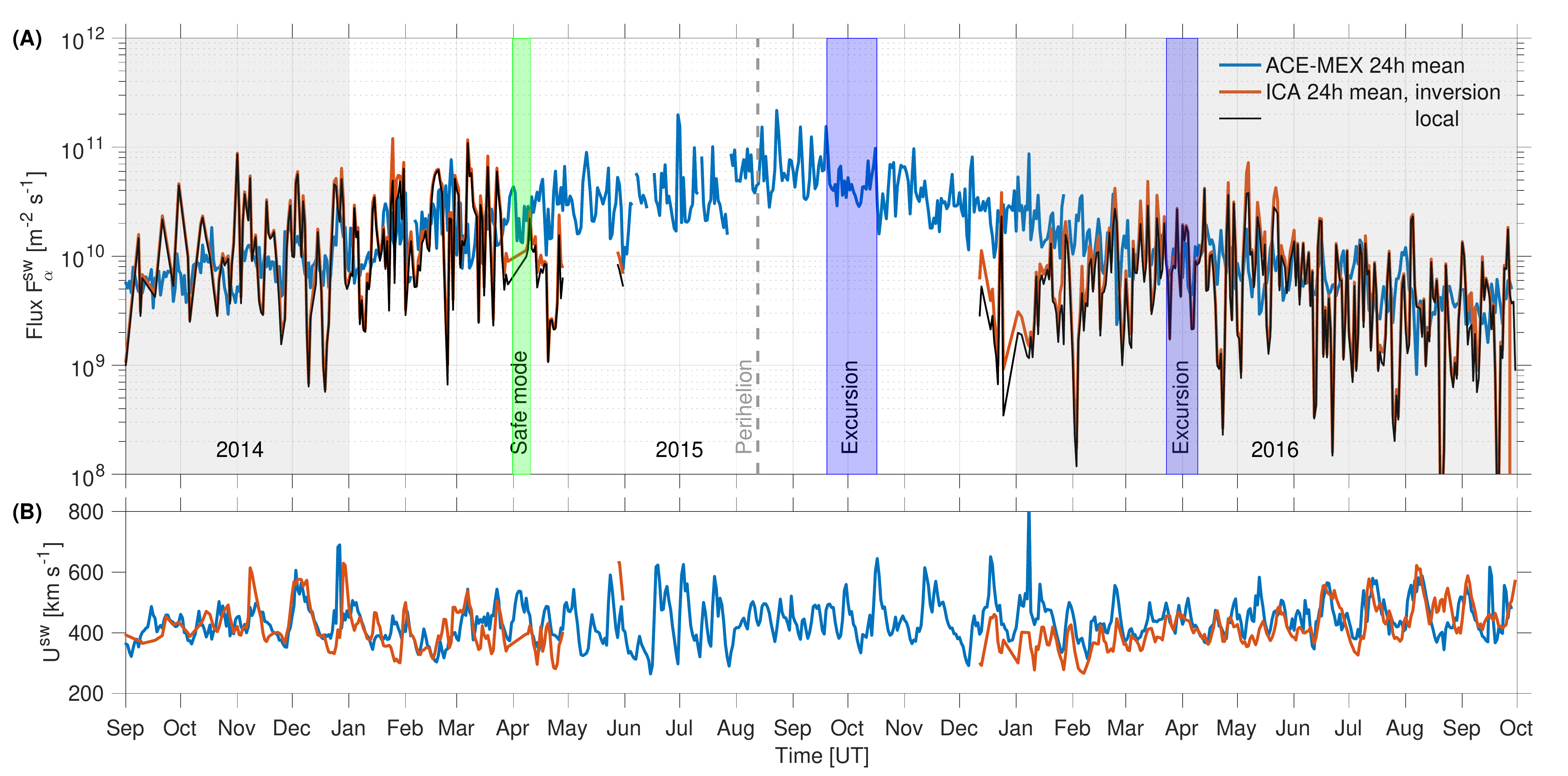}
     \caption{Upstream 'undisturbed' and locally measured solar wind parameters during the \emph{Rosetta} mission (2014-2016), averaged over $\text{one}$ day. (A) Solar wind He$^{2+}$ flux, noted $F^\textnormal{sw}_\alpha$, from ACE-MEX (blue line), and retrieved from RPC-ICA local measurements at comet 67P (red line). Local RPC-ICA fluxes are shown in black for comparison. (B) Solar wind speed measured by ACE-Mars Express (H$^{+}$, blue line) and by RPC-ICA (He$^{2+}$, red line). 
     }
     \label{fig:SWretrieval}
 \end{figure*}

Figure~\ref{fig:SWretrieval}A presents the results of the inversion for He$^{2+}$ ions measured by RPC-ICA in the inner coma. We compared our retrieved solar wind He$^{2+}$ particle flux, $F^\textnormal{sw}_\alpha$ with measurements made at Earth (by the satellites ACE an WIND) and at Mars (by the satellite Mars Express, or MEX), and time- and angle-propagated to the position of comet 67P, assuming a simple Parker-spiral ballistic propagation model \citep[for details, see][]{Behar2018thesis}. The ion spectrometer ASPERA-3 IMA on board MEX \citep{Barabash2006} that we used here was built by the same institute as that of RPC-ICA: thus, both spectrometers are almost identical. ACE measurements (from the Solar Wind Electron, Proton, and Alpha Monitor, SWEPAM electrostatic analyzer) were obtained from the ACE Science Center \citep{McComas1998}. In the figure, we plot the daily average for clarity, combining MEX and ACE data. We conservatively assumed that the solar wind contained about $4\%$ of He$^{2+}$ ions \citep{Slavin1981}, and hence the proton flux $F^\textnormal{sw}_p$ measured by ACE and MEX is $F^\textnormal{sw}_\alpha = \sfrac{1}{24}\,F^\textnormal{sw}_p$.  

Despite strong fluctuations in the cometary data, the daily averages from RPC-ICA and MEX-ACE are in relatively good agreement throughout the mission, until about April 2015, and after February 2016. As pointed out earlier, a clear decrease in flux is seen at these times in the local RPC-ICA data (black line in Fig.~\ref{fig:SWretrieval}). According to our model calculations, this cannot be explained by an increased efficiency of charge-changing reactions: the retrieved upstream particle flux (red line) is only marginally higher than the local flux during these periods, with an increase of a factor of $2$ on average, which is far different from the expected fluxes from MEX-ACE. We note, however, that this efficiency level of charge-changing reactions is in agreement with the conclusions reached in Paper~II for a synthetic test case applied to \emph{Rosetta}: SWCX effects were mostly expected around perihelion, when the neutral density reached sufficiently high levels (see also Appendix~B of Paper~II on charge-exchange collision depth). 

The discrepancy may stem from at least two immediate issues that we describe below. 
\begin{itemize}
    \item Instrumental bias. The RPC-ICA field of view is only about $2.8\pi$~sr, therefore the instrument may have missed several detections of ions, which may have led to an underestimation of the local solar wind ion flux. This in turn would lead in the model to an underestimation of the upstream solar wind flux, especially since the latter is derived from He$^{2+}$ local fluxes alone. When we used He$^+$ ions to derive the upstream solar wind He$^{2+}$ flux, differences of up to a factor $2-5$ between the two retrievals were obtained, which points to the problem of defining the initial conditions in the model. When the outgassing rate was derived (previous section), this problem was circumvented by the use of a ratio of ion fluxes.
    \item Model bias. In the model, the dynamics of solar wind ions is not self-consistently taken into account: for instance, their cycloidal paths and observed deflection \citep{Behar2017} may substantially increase the path of the ions in the atmosphere, and hence the total column density traversed. When we multiply by $4$ the path traversed by the ions, as may be expected from half an arc length in a typical cycloidal motion, we witness a dramatic increase in the retrieved ion flux below $2.5$\,AU that  reaches up to a factor $10$ and matches the levels expected from MEX-ACE measurements.
\end{itemize} 

Moreover, the model does not assume any formation of boundaries upstream of the nucleus, contrary to observational evidence \citep{Gunell2018}. Following Paper~II, we investigated the effect of high Maxwellian temperatures, that is, $T~=~40$\,MK ($\varv_\textnormal{th}\,=\,500$\,\kms{}), on the retrieval of upstream solar wind fluxes. Differences of less than $10\%$ between the two retrieved fluxes were found, in agreement with the findings of Paper~II, which is marginal in comparison to geometrical effects (e.g., the path length). In Fig.~\ref{fig:SWretrieval}B the solar wind ions measured by RPC-ICA do not display any significant slowing-down throughout the mission: they are in good agreement with ACE/MEX upstream measurements, with only an indication of a $50$\,\kms{} deceleration from December 2015 to February 2016. This decrease in speed is small enough to affect the cross section values  very little (see Paper~I).
Consequently, the current balance between all cross sections is expected to remain mostly unaffected. For example, an increase in double-electron capture cross sections for He$^{2+}$ would be expected if solar wind single particle speeds were to drop below $200$\,\kms{}, which was never the case during the mission. 

The main discrepancy between upstream-retrieved and \emph{Rosetta}-propagated fluxes requires further investigation at this stage. It would also benefit from the use of self-consistent models such as hybrid models \citep{Koenders2016,CSW2017}. This is left for a future study.

\subsection{Ionization processes at comet 67P}
\cite{Heritier2018} compared photoionization (PI) frequencies with electron-impact ionization (EI) frequencies, showing that EI played a major role at large heliocentric distances ($>2.8$\,AU) in the creation of new plasma.
Although charge-transfer reactions may not always result in the net creation of ions, they contribute nonetheless to the general ionization of the coma by adding heavy cometary ions to the cometary plasma. Moreover, solar wind ions are also capable of ionizing the neutral cometary atmosphere to produce new ions due to their high kinetic energies.

\begin{figure*}
  \includegraphics[width=\linewidth]{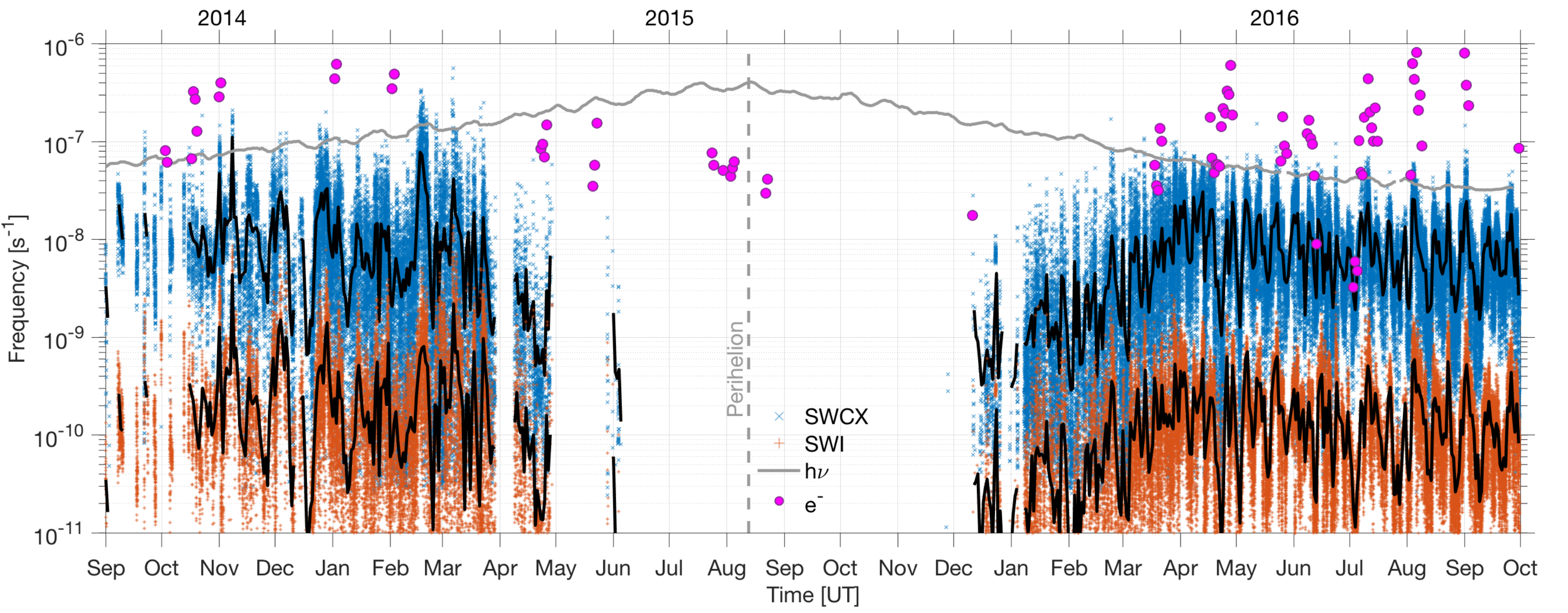}
     \caption{Local ionization production frequencies at comet 67P during the \emph{Rosetta} mission (2014-2016). The contribution of SWCX (blue line), SWI (red lines), PI or $(h\nu$, $24$h average, gray line), and EI (or $e^-$, $24$h average, pink circles) to the local production of ions is displayed. The black continuous lines are the $24$h averaged SWCX and SWI frequencies. When \emph{Rosetta} was inside the solar wind ion cavity (May-December 2015), no solar wind fluxes could be routinely measured.
     }
     \label{fig:ionisationFrequency}
 \end{figure*}

We present in this section the in situ ionization frequency due to SWCX and solar wind impact ionization (SWI), and compare them to PI and EI rates. The results of \cite{Heritier2018} (see their Fig.~$16$) for the PI and EI frequencies are reproduced in Fig.~\ref{fig:ionisationFrequency} with a brief summary below. PI frequencies for an H$_2$O coma at the location of \emph{Rosetta} were estimated from the daily EUV solar fluxes measured at Earth by the Thermosphere Ionosphere Mesosphere Energetics and Dynamics-Solar EUV experiment \citep[TIMED-SEE,][]{Woods2005}, phase- and time-shifted to the position of comet 67P and scaled to its heliocentric distance ($\propto 1/R_\textnormal{Sun}^2$). EI frequencies were calculated using electron fluxes above about $12.5$\,eV from the \emph{Rosetta} Ion and Electron Sensor \citep[RPC-IES,][]{burch2007ssr}. Daily standard deviations are given in \cite{Heritier2018}.

 The SWCX and SWI processes were calculated locally from RPC-ICA solar wind ion particle fluxes $F_i$ (units of m$^{-2}$~s$^{-1}$), with subscript $i=p$, $i=\alpha$, and $i=\textnormal{He}^+$ for H$^+$, He$^{2+}$ , and He$^+$, respectively. The total SWCX frequencies $\sum_i f^\textnormal{swcx}_i$ expressed in s$^{-1}$ are thus
\begin{align}
    f^\textnormal{swcx}_\textnormal{tot} &= f^\textnormal{swcx}_p + f^\textnormal{swcx}_\alpha + f^\textnormal{swcx}_{\textnormal{He}^+},\\
    \textnormal{with}&\nonumber\\
    &\begin{cases}
    f^\textnormal{swcx}_p= \left(\sigma_{10}(U_p)+\sigma_{1-1}(U_p)\right)\ F_p\nonumber\\
    f^\textnormal{swcx}_\alpha= \left(\sigma_{21}(U_\alpha)+\sigma_{20}(U_\alpha)\right)\ F_\alpha\quad .\nonumber\\
    f^\textnormal{swcx}_{\textnormal{He}^{+}}= \sigma_{10}(U_{\textnormal{He}^{+}})\ F_{\textnormal{He}^{+}}\nonumber
    \end{cases}
\end{align}

Hydrogen and helium cross sections $\sigma_{ij}$ were taken at the ion speed $U_i$ measured by RPC-ICA.
Similarly, the total solar wind impact ionization frequencies $\sum_i f^\textnormal{swi}_i$ in s$^{-1}$ were calculated as
\begin{align}
f^\textnormal{swi}_\textnormal{tot} &= f^\textnormal{swi}_p + f^\textnormal{swi}_\alpha + f^\textnormal{swi}_{\textnormal{He}^+},\\
\textnormal{with}&\nonumber\\
    &\begin{cases}
    f^\textnormal{swi}_p= \sigma_{11}(U_p)\ F_p\nonumber\\
    f^\textnormal{swi}_\alpha= \sigma_{22}(U_\alpha)\ F_\alpha\quad .\nonumber\\
    f^\textnormal{swi}_{\textnormal{He}^{+}}= \sigma_{11}(U_{\textnormal{He}^{+}})\ F_{\textnormal{He}^{+}}\nonumber
    \end{cases}
\end{align}

For ionization by protons and $\alpha$ particles, the average energy of the ejected electrons at solar wind energies is expected to be about $5$\,eV \citep{Uehara2000,Uehara2002}.

Figure~\ref{fig:ionisationFrequency} presents a comparison of all ionization channels at comet 67P, assuming only water as neutral constituent of the coma: daily averaged PI (gray line) and EI (pink circles), as well as non-averaged SWCX (blue) and SWI (red). Daily averaged SWCX and SWI frequencies are plotted as black lines overlaid on the non-averaged data; this demonstrates the extreme variability of fluxes measured at \emph{Rosetta} during the mission. During the early mission and toward its end, electron ionization often becomes the largest contribution \citep{Galand2016,Heritier2018} that even exceeds PI, whereas PI remains dominant in the so-called solar wind ion cavity between May and December 2015.

On average, SWCX is a factor $5$ lower than PI at large heliocentric distances (early and late mission, coinciding with cometocentric distances $<50$\,km, see Fig.~\ref{fig:outgassingrate}B). After January 2015 ($R_\textnormal{Sun}<2.8$\,AU), this factor reaches $10$ on average, until the substantial drop in solar wind flux after March 2015 ($R_\textnormal{Sun}<2.4$\,AU), where SWCX becomes $100$ times less efficient than PI. A similar trend is seen for the outbound leg, with an average factor $5$ between PI and SWCX reached after April 2016 ($R_\textnormal{Sun}>2.8$\,AU). This heliocentric distance coincides on the inbound and outbound legs with the sudden drop in solar wind ion fluxes, a prelude to the formation of the solar wind ion cavity \citep{Behar2017}.

Occasionally, SWCX ionization frequencies can become higher than PI frequencies and rival EI rates, as is the case in November 2014, at the end of February 2015, and frequently toward the end of the mission. This is the result of increased fluxes that may be related to solar wind transient events such as interplanetary coronal mass ejections (ICMEs) or corotating interaction regions (CIRs). The October-November 2014 SWCX frequency increases coincide for example with the occurrence of several CIRs, as described in \cite{Edberg2016JGR}. Similarly, \cite{Hajra2018} studied CIR events between June and September 2016: their impact at the comet correlated relatively well with sudden increases in charge-exchange rates; the authors used the same calculation method as presented here. Because of the large daily variations in the flux, however, case-by-case studies must be performed to study in more detail how SWCX rates compare with EI. 

Throughout the mission, SWI is a constant $50$ times lower than SWCX, which makes it a negligible plasma source in comparison to ionization by electron impact and solar EUV radiation. This is stemming from ionization cross sections, which peak at velocities above $1000$\,\kms{}, and are thus small in the typical solar wind velocity ranges (see Paper~I for details). However, this assessment may change if the solar wind is substantially heated, as would be expected with the formation of boundaries and bow shock-like structures upstream. For example, as shown in Paper~I, total proton ionization cross sections may be multiplied by about a factor $10$  at $400$\,\kms{} solar wind speed: such an increase in SWI cross sections would significantly narrow the gap with the SWCX frequency. However, in order to match the levels of PI, an even larger increase in cross section would be necessary (e.g., a factor $100-500$), which may occur for a significant slowing-down of the solar wind (below $200$\,\kms{}). Although this condition is not met at comet 67P during the \emph{Rosetta} mission, such a large deceleration is expected at high-activity comets such as comet 1P/Halley.

\subsection{ENA environment: predictions from RPC-ICA data}
From SWCX reactions, ENAs such as He$^{0}$ and H$^{0}$ can be efficiently produced. Our analytical model of charge-changing processes at the comet is able to reproduce their distribution throughout the \emph{Rosetta} mission (see also Paper~II) using He$^{+}$-to-He$^{2+}$ particle flux ratios to calibrate our model to the observations. For inputs, we used here cross sections at the mean speed of the solar wind measured by RPC-ICA, the column density derived from ROSINA, and, as previously, the neutral outgassing velocity from the \cite{Hansen2016} empirical formula.

\begin{figure*}
  \includegraphics[width=\linewidth]{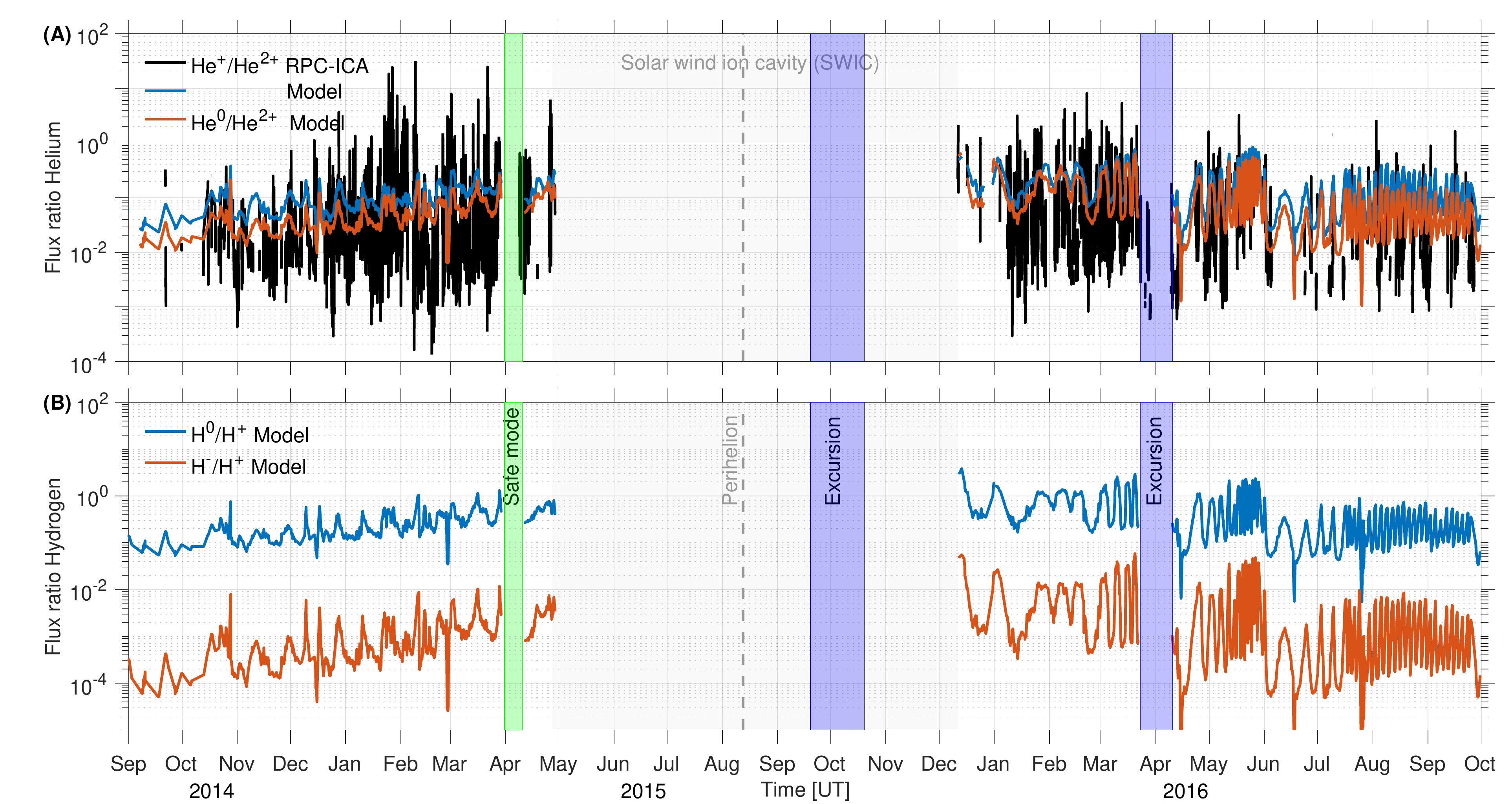}
     \caption{Particle flux ratios during the \emph{Rosetta} mission 2014-2016. (A) Helium species. The 1h averaged He$^+$/He$^{2+}$ ratio measured by RPC-ICA (black) is compared to the daily averaged analytical forward model solution (blue), with mean speed $U_\textnormal{sw} = U_\textnormal{ICA}$(He$^{2+}$). The column density is derived from ROSINA data. Modeled He$^0$/He$^{2+}$ flux ratios are shown in red. (B) Hydrogen species, with modeled H$^0$/H$^{+}$ (blue) and H$^-$/H$^{+}$ (red) ratios for a solar wind mean speed $U_\textnormal{sw} = U_\textnormal{ICA}$(H$^{+}$). 
     The solar wind ion cavity is indicated as a gradually denser gray-shaded region. 
     Safe mode and excursions where ROSINA data were excluded from the analysis are indicated.}
     \label{fig:ratiosMission}
 \end{figure*}

Figure~\ref{fig:ratiosMission}A shows first the reconstructed in situ He$^+$/He$^{2+}$ particle flux ratios (in blue, noted $\mathcal{R}_{\textnormal{He}^+}$ in the following) as compared to those measured locally by RPC-ICA (in black). Excellent agreement between the two curves is achieved throughout the mission, which \textup{\emph{\textup{a posteriori}}} is another confirmation of the accuracy of the outgassing retrieval performed in Section\,\ref{sec:outgassingRetrieval} when the RPC-ICA flux ratio is used to estimate it. This enables us to derive with confidence the corresponding local flux ratio for the production of helium ENAs He$^{0}$/He$^{2+}$, noted $\mathcal{R}_{\textnormal{He}}$. At the beginning of the mission, the ENA He$^0$ flux at \emph{Rosetta} is expected to be about a factor $2$ lower than that of He$^+$. When the He$^+$ flux starts to drop in March 2015 (see Fig.~\ref{fig:ICAmoments}), this factor is closer to one, meaning that in the denser atmosphere of this time period, there would be as many He$^+$ ions than He ENAs at \emph{Rosetta}'s location. The tendency is identical after perihelion for the outbound leg, where ENAs and He$^+$ ions are in the same proportion in the solar wind plasma until March 2016. After this, He$^+$ ion fluxes routinely become $2-3$ times as high as He ENAs. In extremely rare events, we predict that ENAs can come to $50\%$ of He$^{2+}$ and dominate over He$^+$ ions: this occurs once on 24 April 2015, and after perihelion on 12 and 31 December 2015, and on 6-7 February 2016. A detailed study of such events is beyond the scope of this paper, but may be tentatively linked to sudden changes in the upstream solar wind plasma, driven by solar transients such as CIRs \citep{Hajra2018}.

We calculated the ENA ratio for hydrogen and show it in Fig.~\ref{fig:ratiosMission}B. Because of high proton fluxes outside of the solar wind ion cavity, the proportion of H ENAs is expectedly large throughout the mission; the ENA flux may even become higher than the proton flux in very rare events (ratio above $1$, as in February and March 2015, and after perihelion in December 2015, occasionally in January and February 2016, and large periods of time in March and May 2016). The highest predicted ENA fluxes are immediately after \emph{Rosetta} exited the SWIC. Correspondingly, the proportion of H$^-$ negative ions is predicted to be no more than $0.01\%$ of the proton flux throughout most of the mission, reaching higher levels in the same transient events as for H ENAs.

These ratios can in turn be used to predict the total fluxes of ENAs that could have been measured had \emph{Rosetta} included an ENA detector. For helium ENAs,
\begin{align}
    F_\textnormal{He} &= \mathcal{R}_{\textnormal{He}}\ F_{\alpha}^\textnormal{ica} = \frac{\mathcal{R}_{\textnormal{He}}}{\mathcal{R}_{\textnormal{He}^+}^\textnormal{ica}} F_{\textnormal{He}^+}^\textnormal{ica},\label{eq:HeENA}\\
    &\textnormal{with}\quad
        \mathcal{R}_{\textnormal{He}}= \frac{F_\textnormal{He}}{F_{\alpha}} \quad\textnormal{and}\quad \mathcal{R}_{\textnormal{He}^+}^\textnormal{ica}= \frac{F_{\textnormal{He}^+}}{F_{\alpha}}.\nonumber
\end{align}

Ion fluxes for He$^{2+}$ and He$^+$, noted $F_\alpha^\textnormal{ica}$ and $F_{\textnormal{He}^+}^\textnormal{ica}$, and the particle flux ratio $\mathcal{R}_{\textnormal{He}^+}^\textnormal{ica}$ are here measured by RPC-ICA, whereas the ratio $\mathcal{R}_{\textnormal{He}}$ is calculated in Fig.~\ref{fig:ratiosMission} by the analytical model, using ROSINA-COPS data for the neutral atmosphere. To better constrain the model with the flux ratio $\mathcal{R}_{\textnormal{He}^+}$ measured by RPC-ICA (in an attempt to ignore fluctuating upstream solar wind conditions), we chose the second expression in Eq.~(\ref{eq:HeENA}), which multiplies the ratio of flux ratios by the measured He$^+$ flux, acting as a calibration factor for the model.

For hydrogen ENAs and negative ions H$^-$, a similar development yields
\begin{align}
    F_\textnormal{H} &= \mathcal{R}_{\textnormal{H}}\ F_p^\textnormal{ica},\\
    F_{\textnormal{H}^-} &= \mathcal{R}_{\textnormal{H}^-}\ F_p^\textnormal{ica},\\
    \textnormal{with}&\quad \mathcal{R}_{\textnormal{H}}= \frac{F_\textnormal{H}}{F_p} \quad\textnormal{and}\quad \mathcal{R}_{\textnormal{H}^-} = \frac{F_{\textnormal{H}^-}}{F_p}, \nonumber
\end{align}
with $F_p^\textnormal{ica}$ the flux of protons measured by RPC-ICA. All other quantities were calculated by the model.

\begin{figure*}
  \includegraphics[width=\linewidth]{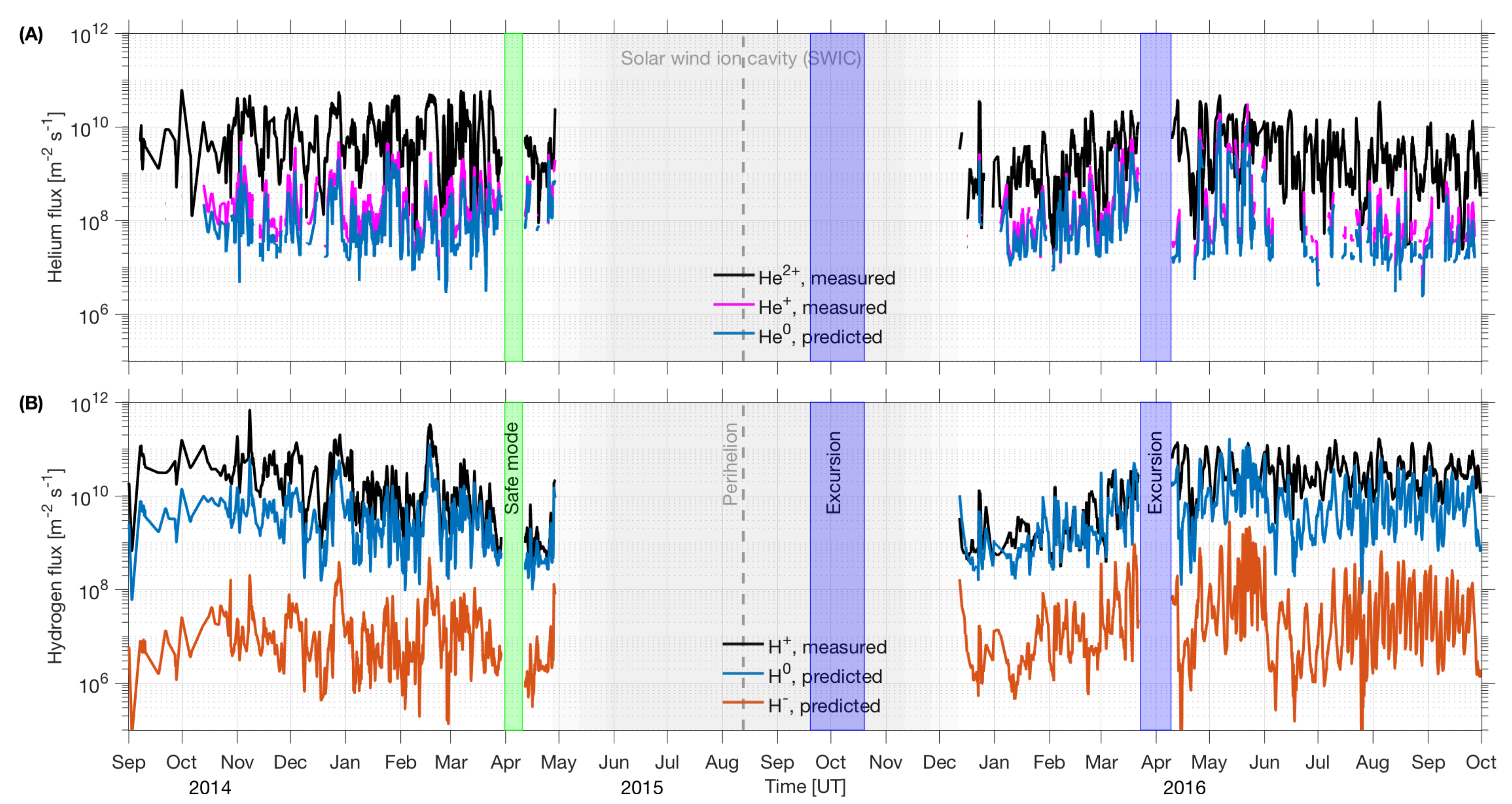}
     \caption{Solar wind ion and ENA fluxes during the \emph{Rosetta} mission 2014-2016. The fluxes are measured when available by RPC-ICA, and when unavailable, are predicted by the analytical model using ROSINA-COPS and RPC-ICA data. All fluxes are averaged over $24$~h. (A) Helium species. (B) Hydrogen species. ENA fluxes are drawn in blue for clarity. The solar wind ion cavity is indicated as a gray-shaded region.
     Safe mode and excursions where ROSINA data were excluded from the analysis are indicated.}
     \label{fig:ENAfluxesMission}
 \end{figure*}

Figures~\ref{fig:ENAfluxesMission}A and B present the synoptic summary of daily averaged fluxes of solar wind origin at comet 67P at the position of \emph{Rosetta}. He$^{2+}$, He$^{+}$, and H$^{+}$ fluxes are all measured by RPC-ICA, whereas all other particles fluxes are predicted by the model. We note, as before, that He$^+$ and He ENAs have similar levels throughout the mission, with occasional spikes in magnitude that propel them to the levels of He$^{2+}$ ions. This suggests that most of the time, \emph{Rosetta} orbited at a cometocentric distance where He$^+$ and He$^0$ charge states were equally distributed (see Paper~II for cometocentric profiles).

Regarding hydrogen particles, it is clear from Fig.~\ref{fig:ENAfluxesMission}B that H ENAs and protons have similar flux levels throughout the mission, except at its very beginning, and, marginally, at its very end. Possible effects of an increased ENA environment are discussed in \cite{Ip1990} for comet 1P/Halley and include heating of the cometary ions. H$^-$ anions have very low fluxes throughout the mission; despite this, \cite{Burch2015} found evidence of H$^-$ in the early RPC-IES datasets, which persisted up to January 2015. Our model predicts the most favorable detection conditions for H$^-$ in December 2015, around mid-March 2016 and in May 2016. This would need to be investigated further with RPC-IES: flux ratios could give an indication of charge-exchange efficiencies, which could easily be compared with our model to RPC-IES and RPC-ICA datasets for helium.

%

\section{Conclusions}
This study is the culmination of our investigation of charge-changing reactions in cometary atmospheres. \cite{CSW2018a} (Paper~I) made a careful survey of available velocity-dependent charge-changing and ionization cross sections in H$_2$O. \cite{CSW2018b} (Paper~II) developed a new analytical model of charge-changing reactions in comets based on these cross sections, with a systematic exploration of the parameter space (heliocentric and cometocentric distances) of our simulations. In this study, we have applied this model to the complex datasets of the cornerstone ESA-\emph{Rosetta} mission to comet 67P/Churyumov-Gerasimenko. We have provided mission-wide interpretations of the RPC-ICA ion spectrometer data and attempted to place them in the larger context of cometary plasma physics. More specifically, we have shown the following. 
   \begin{enumerate}
      \item Single-electron and double-electron capture cross sections dominate for H$^+$ and He$^{2+}$ solar wind ions interacting with a H$_2$O 67P-like coma. Electron stripping may have played a role at 67P for large heliocentric and small cometocentric distances.
      \item Remote sensing of the cometary neutral atmosphere from local ion measurements is possible. Outgassing rates derived from the local He$^+$/He$^{2+}$ flux ratio were well within a factor $2$ of the neutral pressure sensor estimates: 14-day averages between January 2015 and June 2016 were within $10\%$ of each other and the MIRO estimates. Our fitted production rates agree well with those of \cite{Hansen2016}.
      \item Solar wind upstream retrievals from local ion measurements are difficult, especially for high solar wind ion deflections when closing in on the SWIC. The model was not able to explain the sharp drop in flux in this region, which may either indicate that other mechanisms are at work or that our current approach has problems in taking charge exchange into account. The latter may stem from limiting assumptions in the analytical model (e.g., it does not take the dynamics of individual ions into account), and  possibly from RPC-ICA missing detecting ions in the pre-SWIC region. It is not known at this stage if an additional constraint on the observed downstream fluxes would yield a better estimate in the model.
      \item Charge-exchange reactions play an important role in the inner coma of comet 67P. A comparative summary of all ionization processes, PI, EI, SWCX, and SWI during the \emph{Rosetta} mission was presented. We identified periods when SWCX may for a short time rival electron ionization frequencies in the production of ions. Throughout the mission, SWI played only a minor role, except when significant solar wind heating and strong deceleration of the solar wind flow were concomitantly present; these conditions may be best simultaneously fulfilled for high-activity comets such as comet 1P/Halley.
      \item Hydrogen and helium ENAs are expected to play an important role in the inner coma, with modeled fluxes predicted to match and occasionally even exceed the levels of protons and $\alpha$ particles during the \emph{Rosetta} mission. Enhanced ENA fluxes may in turn lead to localized heating in the coma.
   \end{enumerate}

Our approach may be improved in two main directions. On the one hand, the model inputs could be better constrained: a better determination of charge-changing cross sections or a complementary use of truly 3D ion and ENA measurements to determine the flux ratios would help decrease the errors in the retrievals. On the other hand, improvements on the physics of the model are clearly possible, but would come at the expense of the simplicity and portability of our initial approach: they include the addition of a realistic solar wind ion dynamics (gyromotion and deflection), or that of a more realistic neutral atmosphere (asymmetric outgassing), for example.

As it stands, our analytical model may readily be coupled with test particle simulations using electromagnetic fields calculated by hybrid models to investigate the charge-state distribution of ions in a cometary atmosphere. As a simple tool to interpret usually complex datasets, it can also deliver charge-exchange diagnoses in a wide variety of environments, ranging from astrophysical environments (interstellar medium, etc.) to planetary atmospheres.

\begin{acknowledgements}
The work at the University of Oslo was funded by the Norwegian Research Council grant No. 240000. Work at the Royal Belgian Institute for Space Aeronomy was supported by the Belgian Science Policy Office through the Solar-Terrestrial Centre of Excellence. Work at Ume\aa{} University was funded by SNSB grant 201/15 and SNSA grant 108/18. The work at NASA/SSAI was supported by NASA Astrobiology Institute grant NNX15AE05G and by the NASA HIDEE Program. 
Work at Imperial College London was supported by STFC of UK under grant ST/N000692/1 and by ESA under contract No. 4000119035/16/ES/JD.
The authors thank the ISSI International Team "Plasma Environment of comet 67P after \emph{Rosetta}" for fruitful discussions and collaborations. C.S.W. thanks M.S.W. for inspiring discussions and ideas to improve the manuscript and figures. 
Datasets of the \emph{Rosetta} mission can be freely accessed from ESA's Planetary Science Archive (\url{http://archives.esac.esa.int/psa}).
\end{acknowledgements}


\bibliographystyle{aa}
\bibliography{references_all} 

\begin{thebibliography}{57}
\expandafter\ifx\csname natexlab\endcsname\relax\def\natexlab#1{#1}\fi

\bibitem[{{Alho} {et~al.}(2019){Alho}, {Simon Wedlund}, {Nilsson}, {Kallio},
  {Jarvinen}, \& {Pulkkinen}}]{Alho2019}
{Alho}, M., {Simon Wedlund}, C., {Nilsson}, H., {et~al.} 2019, submitted to
  A\&A, 1, 1

\bibitem[{{Balsiger} {et~al.}(2007){Balsiger}, {Altwegg}, {Bochsler},
  {Eberhardt}, {Fischer}, {Graf}, {J{\"a}ckel}, {Kopp}, {Langer}, {Mildner},
  {M{\"u}ller}, {Riesen}, {Rubin}, {Scherer}, {Wurz}, {W{\"u}thrich}, {Arijs},
  {Delanoye}, {de Keyser}, {Neefs}, {Nevejans}, {R{\`e}me}, {Aoustin},
  {Mazelle}, {M{\'e}dale}, {Sauvaud}, {Berthelier}, {Bertaux}, {Duvet},
  {Illiano}, {Fuselier}, {Ghielmetti}, {Magoncelli}, {Shelley}, {Korth},
  {Heerlein}, {Lauche}, {Livi}, {Loose}, {Mall}, {Wilken}, {Gliem}, {Fiethe},
  {Gombosi}, {Block}, {Carignan}, {Fisk}, {Waite}, {Young}, \&
  {Wollnik}}]{Balsiger2007}
{Balsiger}, H., {Altwegg}, K., {Bochsler}, P., {et~al.} 2007, Space Sci. Rev.,
  128, 745

\bibitem[{{Banks} \& {Kockarts}(1973)}]{Banks1973a}
{Banks}, P.~M. \& {Kockarts}, G. 1973, {Aeronomy, Part A} (Academic Press, New
  York and London)

\bibitem[{{Barabash} {et~al.}(2006){Barabash}, {Lundin}, {Andersson},
  {Brinkfeldt}, {Grigoriev}, {Gunell}, {Holmstr{\"o}m}, {Yamauchi}, {Asamura},
  {Bochsler}, {Wurz}, {Cerulli-Irelli}, {Mura}, {Milillo}, {Maggi}, {Orsini},
  {Coates}, {Linder}, {Kataria}, {Curtis}, {Hsieh}, {Sandel}, {Frahm},
  {Sharber}, {Winningham}, {Grande}, {Kallio}, {Koskinen}, {Riihel{\"a}},
  {Schmidt}, {S{\"a}les}, {Kozyra}, {Krupp}, {Woch}, {Livi}, {Luhmann},
  {McKenna-Lawlor}, {Roelof}, {Williams}, {Sauvaud}, {Fedorov}, \&
  {Thocaven}}]{Barabash2006}
{Barabash}, S., {Lundin}, R., {Andersson}, H., {et~al.} 2006, Space Sci. Rev.,
  126, 113

\bibitem[{{Behar}(2018)}]{Behar2018thesis}
{Behar}, E. 2018, PhD thesis, Lule\aa{} University of Technology, Space
  Technology

\bibitem[{{Behar} {et~al.}(2016{\natexlab{a}}){Behar}, {Lindkvist}, {Nilsson},
  {Holmstr{\"o}m}, {Stenberg-Wieser}, {Ramstad}, \& {G{\"o}tz}}]{Behar2016b}
{Behar}, E., {Lindkvist}, J., {Nilsson}, H., {et~al.} 2016{\natexlab{a}}, A\&A,
  596, A42

\bibitem[{{Behar} {et~al.}(2017){Behar}, {Nilsson}, {Alho}, {Goetz}, \&
  {Tsurutani}}]{Behar2017}
{Behar}, E., {Nilsson}, H., {Alho}, M., {Goetz}, C., \& {Tsurutani}, B. 2017,
  Month. Not. Roy. Astron. Soc., 469, S396

\bibitem[{{Behar} {et~al.}(2018{\natexlab{a}}){Behar}, {Nilsson}, {Henri},
  {Ber{\v c}i{\v c}}, {Nicolaou}, {Stenberg Wieser}, {Wieser}, {Tabone},
  {Saillenfest}, \& {Goetz}}]{Behar2018tail}
{Behar}, E., {Nilsson}, H., {Henri}, P., {et~al.} 2018{\natexlab{a}}, \aap,
  616, A21

\bibitem[{{Behar} {et~al.}(2016{\natexlab{b}}){Behar}, {Nilsson}, {Wieser},
  {Nemeth}, {Broiles}, \& {Richter}}]{Behar2016a}
{Behar}, E., {Nilsson}, H., {Wieser}, G.~S., {et~al.} 2016{\natexlab{b}},
  Geophys. Res. Lett., 43, 1411

\bibitem[{{Behar} {et~al.}(2018{\natexlab{b}}){Behar}, {Tabone}, {Saillenfest},
  {Henri}, {Deca}, {Lindkvist}, {Holmstr{\"o}m}, \&
  {Nilsson}}]{Behar2018aa_model}
{Behar}, E., {Tabone}, B., {Saillenfest}, M., {et~al.} 2018{\natexlab{b}},
  \aap, 620, A35

\bibitem[{{Ber{\v c}i{\v c}} {et~al.}(2018){Ber{\v c}i{\v c}}, {Behar},
  {Nilsson}, {Nicolaou}, {Wieser}, {Wieser}, \& {Goetz}}]{Bercic2018}
{Ber{\v c}i{\v c}}, L., {Behar}, E., {Nilsson}, H., {et~al.} 2018, \aap, 613,
  A57

\bibitem[{{Beth} {et~al.}(2016){Beth}, {Altwegg}, {Balsiger}, {Berthelier},
  {Calmonte}, {Combi}, {De Keyser}, {Dhooghe}, {Fiethe}, {Fuselier}, {Galand},
  {Gasc}, {Gombosi}, {Hansen}, {H{\"a}ssig}, {H{\'e}ritier}, {Kopp}, {Le Roy},
  {Mandt}, {Peroy}, {Rubin}, {S{\'e}mon}, {Tzou}, \& {Vigren}}]{Beth2016}
{Beth}, A., {Altwegg}, K., {Balsiger}, H., {et~al.} 2016, Month. Not. Roy.
  Astron. Soc., 462, S562

\bibitem[{{Bieler} {et~al.}(2015{\natexlab{a}}){Bieler}, {Altwegg}, {Balsiger},
  {Bar-Nun}, {Berthelier}, {Bochsler}, {Briois}, {Calmonte}, {Combi}, {de
  Keyser}, {van Dishoeck}, {Fiethe}, {Fuselier}, {Gasc}, {Gombosi}, {Hansen},
  {H{\"a}ssig}, {J{\"a}ckel}, {Kopp}, {Korth}, {Le Roy}, {Mall}, {Maggiolo},
  {Marty}, {Mousis}, {Owen}, {R{\`e}me}, {Rubin}, {S{\'e}mon}, {Tzou}, {Waite},
  {Walsh}, \& {Wurz}}]{Bieler2015nature}
{Bieler}, A., {Altwegg}, K., {Balsiger}, H., {et~al.} 2015{\natexlab{a}}, \nat,
  526, 678

\bibitem[{{Bieler} {et~al.}(2015{\natexlab{b}}){Bieler}, {Altwegg}, {Balsiger},
  {Berthelier}, {Calmonte}, {Combi}, {De Keyser}, {Fiethe}, {Fougere},
  {Fuselier}, {Gasc}, {Gombosi}, {Hansen}, {H{\"a}ssig}, {Huang}, {J{\"a}ckel},
  {Jia}, {Le Roy}, {Mall}, {R{\`e}me}, {Rubin}, {Tenishev}, {T{\'o}th}, {Tzou},
  \& {Wurz}}]{Bieler2015}
{Bieler}, A., {Altwegg}, K., {Balsiger}, H., {et~al.} 2015{\natexlab{b}}, A\&A,
  583, A7

\bibitem[{{Bodewits} {et~al.}(2006){Bodewits}, {Hoekstra}, {Seredyuk},
  {McCullough}, {Jones}, \& {Tielens}}]{Bodewits2006}
{Bodewits}, D., {Hoekstra}, R., {Seredyuk}, B., {et~al.} 2006, Astrophys. J.,
  642, 593

\bibitem[{Burch {et~al.}(2007)Burch, Goldstein, Cravens, Gibson, Lundin,
  Pollock, Winningham, \& Young}]{burch2007ssr}
Burch, J., Goldstein, R., Cravens, T., {et~al.} 2007, Space Science Reviews,
  128, 697

\bibitem[{{Burch} {et~al.}(2015){Burch}, {Cravens}, {Llera}, {Goldstein},
  {Mokashi}, {Tzou}, \& {Broiles}}]{Burch2015}
{Burch}, J.~L., {Cravens}, T.~E., {Llera}, K., {et~al.} 2015, Geophys. Res.
  Lett., 42, 5125

\bibitem[{{Combi} {et~al.}(2004){Combi}, {Harris}, \& {Smyth}}]{Combi2004}
{Combi}, M.~R., {Harris}, W.~M., \& {Smyth}, W.~H. 2004, in Comets II, ed.
  M.~C. {Festou}, H.~U. {Keller}, \& H.~A. {Weaver} (1510 E. University Blvd.,
  P.O. Box 210055, Tucson, AZ 85721-0055: The University of Arizona Press),
  523--552

\bibitem[{{Cravens} \& {Gombosi}(2004)}]{Cravens2004}
{Cravens}, T.~E. \& {Gombosi}, T.~I. 2004, Adv. Space Res., 33, 1968

\bibitem[{{Dennerl}(2010)}]{Dennerl2010}
{Dennerl}, K. 2010, Space Sci. Rev., 157, 57

\bibitem[{{Edberg} {et~al.}(2016){Edberg}, {Eriksson}, {Odelstad}, {Vigren},
  {Andrews}, {Johansson}, {Burch}, {Carr}, {Cupido}, {Glassmeier}, {Goldstein},
  {Halekas}, {Henri}, {Koenders}, {Mandt}, {Mokashi}, {Nemeth}, {Nilsson},
  {Ramstad}, {Richter}, \& {Wieser}}]{Edberg2016JGR}
{Edberg}, N.~J.~T., {Eriksson}, A.~I., {Odelstad}, E., {et~al.} 2016, J.
  Geophys. Res., 121, 949

\bibitem[{{Fougere} {et~al.}(2016){Fougere}, {Altwegg}, {Berthelier}, {Bieler},
  {Bockel{\'e}e-Morvan}, {Calmonte}, {Capaccioni}, {Combi}, {De Keyser},
  {Debout}, {Erard}, {Fiethe}, {Filacchione}, {Fink}, {Fuselier}, {Gombosi},
  {Hansen}, {H{\"a}ssig}, {Huang}, {Le Roy}, {Leyrat}, {Migliorini},
  {Piccioni}, {Rinaldi}, {Rubin}, {Shou}, {Tenishev}, {Toth}, \&
  {Tzou}}]{Fougere2016}
{Fougere}, N., {Altwegg}, K., {Berthelier}, J.-J., {et~al.} 2016, Month. Not.
  Roy. Astron. Soc., 462, S156

\bibitem[{{Fuselier} {et~al.}(2016){Fuselier}, {Altwegg}, {Balsiger},
  {Berthelier}, {Beth}, {Bieler}, {Briois}, {Broiles}, {Burch}, {Calmonte},
  {Cessateur}, {Combi}, {De Keyser}, {Fiethe}, {Galand}, {Gasc}, {Gombosi},
  {Gunell}, {Hansen}, {H{\"a}ssig}, {Heritier}, {Korth}, {Le Roy},
  {Luspay-Kuti}, {Mall}, {Mandt}, {Petrinec}, {R{\`e}me}, {Rinaldi}, {Rubin},
  {S{\'e}mon}, {Trattner}, {Tzou}, {Vigren}, {Waite}, \& {Wurz}}]{Fuselier2016}
{Fuselier}, S.~A., {Altwegg}, K., {Balsiger}, H., {et~al.} 2016, Month. Not.
  Roy. Astron. Soc., 462, S67

\bibitem[{{Fuselier} {et~al.}(2015){Fuselier}, {Altwegg}, {Balsiger},
  {Berthelier}, {Bieler}, {Briois}, {Broiles}, {Burch}, {Calmonte},
  {Cessateur}, {Combi}, {De Keyser}, {Fiethe}, {Galand}, {Gasc}, {Gombosi},
  {Gunell}, {Hansen}, {H{\"a}ssig}, {J{\"a}ckel}, {Korth}, {Le Roy}, {Mall},
  {Mandt}, {Petrinec}, {Raghuram}, {R{\`e}me}, {Rinaldi}, {Rubin}, {S{\'e}mon},
  {Trattner}, {Tzou}, {Vigren}, {Waite}, \& {Wurz}}]{Fuselier2015}
{Fuselier}, S.~A., {Altwegg}, K., {Balsiger}, H., {et~al.} 2015, \aap, 583, A2

\bibitem[{{Galand} {et~al.}(2016){Galand}, {H{\'e}ritier}, {Odelstad}, {Henri},
  {Broiles}, {Allen}, {Altwegg}, {Beth}, {Burch}, {Carr}, {Cupido}, {Eriksson},
  {Glassmeier}, {Johansson}, {Lebreton}, {Mandt}, {Nilsson}, {Richter},
  {Rubin}, {Sagni{\`e}res}, {Schwartz}, {S{\'e}mon}, {Tzou}, {Valli{\`e}res},
  {Vigren}, \& {Wurz}}]{Galand2016}
{Galand}, M., {H{\'e}ritier}, K.~L., {Odelstad}, E., {et~al.} 2016, Month. Not.
  Roy. Astron. Soc., 462, S331

\bibitem[{{Glassmeier}(2017)}]{Glassmeier2017}
{Glassmeier}, K.-H. 2017, Phil. Trans. Roy. Soc. Lond. Ser. A, 375, 20160256

\bibitem[{Glassmeier {et~al.}(2007)Glassmeier, Boehnhardt, Koschny, K{\"u}hrt,
  \& Richter}]{glassmeier2007ssr}
Glassmeier, K.-H., Boehnhardt, H., Koschny, D., K{\"u}hrt, E., \& Richter, I.
  2007, Space Sci. Rev., 128, 1

\bibitem[{{Gombosi}(1987)}]{Gombosi1987}
{Gombosi}, T.~I. 1987, Geophys. Res. Lett., 14, 1174

\bibitem[{{Gunell} {et~al.}(2018){Gunell}, {Goetz}, {Simon Wedlund},
  {Lindkvist}, {Hamrin}, {Nilsson}, {Llera}, {Eriksson}, \&
  {Holmstr{\"o}m}}]{Gunell2018}
{Gunell}, H., {Goetz}, C., {Simon Wedlund}, C., {et~al.} 2018, \aap, 619, L2

\bibitem[{{Hajra} {et~al.}(2018){Hajra}, {Henri}, {Myllys}, {H{\'e}ritier},
  {Galand}, {Simon Wedlund}, {Breuillard}, {Behar}, {Edberg}, {Goetz},
  {Nilsson}, {Eriksson}, {Goldstein}, {Tsurutani}, {Mor{\'e}}, {Valli{\`e}res},
  \& {Wattieaux}}]{Hajra2018}
{Hajra}, R., {Henri}, P., {Myllys}, M., {et~al.} 2018, Month. Not. Roy. Astron.
  Soc., 480, 4544

\bibitem[{{Hansen} {et~al.}(2016){Hansen}, {Altwegg}, \& {et al.}}]{Hansen2016}
{Hansen}, K.~C., {Altwegg}, K., \& {et al.} 2016, Month. Not. Roy. Astr. Soc.,
  1, 1

\bibitem[{{Haser}(1957)}]{Haser1957}
{Haser}, L. 1957, Bull. Soc. Roy. Scie. Li{\`e}ge, 43, 740

\bibitem[{{Heritier} {et~al.}(2018){Heritier}, {Galand}, {Henri}, {Johansson},
  {Beth}, {Eriksson}, {Vallières}, {Altwegg}, {Burch}, {Carr}, {Ducrot},
  {Hajra}, \& {Rubin}}]{Heritier2018}
{Heritier}, K., {Galand}, M., {Henri}, P., {et~al.} 2018, A\&A, 617, 1

\bibitem[{{Huddleston} {et~al.}(1998){Huddleston}, {Strangeway}, {Warnecke},
  {Russell}, \& {Kivelson}}]{Huddleston1998}
{Huddleston}, D.~E., {Strangeway}, R.~J., {Warnecke}, J., {Russell}, C.~T., \&
  {Kivelson}, M.~G. 1998, \jgr, 103, 19887

\bibitem[{{Ip}(1990)}]{Ip1990}
{Ip}, W.-H. 1990, \apj, 353, 290

\bibitem[{{Jones} {et~al.}(2017){Jones}, {Knight}, {Fitzsimmons}, \&
  {Taylor}}]{Jones2017}
{Jones}, G.~H., {Knight}, M.~M., {Fitzsimmons}, A., \& {Taylor}, M.~G.~G.~T.
  2017, Phil. Trans. Roy. Soc. London Ser. A, 375, 20170001

\bibitem[{{Koenders} {et~al.}(2016){Koenders}, {Goetz}, {Richter},
  {Motschmann}, \& {Glassmeier}}]{Koenders2016}
{Koenders}, C., {Goetz}, C., {Richter}, I., {Motschmann}, U., \& {Glassmeier},
  K.-H. 2016, Month. Not. Roy. Astron. Soc., 462, S235

\bibitem[{{L\"{a}uter} {et~al.}(2019){L\"{a}uter}, {Kramer}, {Rubin}, \&
  {Altwegg}}]{Lauter2018}
{L\"{a}uter}, M., {Kramer}, T., {Rubin}, M., \& {Altwegg}, K. 2019, Month. Not.
  Roy. Astron. Soc., 483, 852

\bibitem[{{Marshall} {et~al.}(2017){Marshall}, {Hartogh}, {Rezac}, {von
  Allmen}, {Biver}, {Bockel{\'e}e-Morvan}, {Crovisier}, {Encrenaz}, {Gulkis},
  {Hofstadter}, {Ip}, {Jarchow}, {Lee}, \& {Lellouch}}]{Marshall2017}
{Marshall}, D.~W., {Hartogh}, P., {Rezac}, L., {et~al.} 2017, \aap, 603, A87

\bibitem[{{McComas} {et~al.}(1998){McComas}, {Bame}, {Barker}, {Feldman},
  {Phillips}, {Riley}, \& {Griffee}}]{McComas1998}
{McComas}, D.~J., {Bame}, S.~J., {Barker}, P., {et~al.} 1998, \ssr, 86, 563

\bibitem[{{Nilsson} {et~al.}(2007){Nilsson}, Lundin, Lundin, Barabash, Borg,
  Norberg, Fedorov, Sauvaud, Koskinen, Kallio, Riihel{\"a}, \&
  Burch}]{Nilsson2007}
{Nilsson}, H., Lundin, R., Lundin, K., {et~al.} 2007, Space Science Reviews,
  128, 671

\bibitem[{{Nilsson} {et~al.}(2015{\natexlab{a}}){Nilsson}, {Stenberg Wieser},
  {Behar}, {Simon Wedlund}, {Gunell}, {Yamauchi}, {Lundin}, {Barabash},
  {Wieser}, {Carr}, {Cupido}, {Burch}, {Fedorov}, {Sauvaud}, {Koskinen},
  {Kallio}, {Lebreton}, {Eriksson}, {Edberg}, {Goldstein}, {Henri}, {Koenders},
  {Mokashi}, {Nemeth}, {Richter}, {Szego}, {Volwerk}, {Vallat}, \&
  {Rubin}}]{Nilsson2015}
{Nilsson}, H., {Stenberg Wieser}, G., {Behar}, E., {et~al.} 2015{\natexlab{a}},
  Science, 347, 571

\bibitem[{{Nilsson} {et~al.}(2015{\natexlab{b}}){Nilsson}, {Stenberg Wieser},
  {Behar}, {Simon Wedlund}, {Kallio}, {Gunell}, {Edberg}, {Eriksson},
  {Yamauchi}, {Koenders}, {Wieser}, {Lundin}, {Barabash}, {Mandt}, {Burch},
  {Goldstein}, {Mokashi}, {Carr}, {Cupido}, {Fox}, {Szego}, {Nemeth},
  {Fedorov}, {Sauvaud}, {Koskinen}, {Richter}, {Lebreton}, {Henri}, {Volwerk},
  {Vallat}, \& {Geiger}}]{Nilsson2015AA}
{Nilsson}, H., {Stenberg Wieser}, G., {Behar}, E., {et~al.} 2015{\natexlab{b}},
  A\&A, 583, A20

\bibitem[{{Nilsson} {et~al.}(2017){Nilsson}, {Wieser}, {Behar}, {Gunell},
  {Wieser}, {Galand}, {Simon Wedlund}, {Alho}, {Goetz}, {Yamauchi}, {Henri},
  {Odelstad}, \& {Vigren}}]{Nilsson2017a}
{Nilsson}, H., {Wieser}, G.~S., {Behar}, E., {et~al.} 2017, Month. Not. Roy.
  Astron. Soc., 469, S252

\bibitem[{{Simon Wedlund} {et~al.}(2017){Simon Wedlund}, {Alho}, {Gronoff},
  {Kallio}, {Gunell}, {Nilsson}, {Lindkvist}, {Behar}, {Stenberg Wieser}, \&
  {Miloch}}]{CSW2017}
{Simon Wedlund}, C., {Alho}, M., {Gronoff}, G., {et~al.} 2017, A\&A, 604, A73

\bibitem[{{Simon Wedlund} {et~al.}(2019{\natexlab{a}}){Simon Wedlund}, {Behar},
  {Kallio}, {Nilsson}, {Alho}, {Gunell}, {Bodewits}, {Beth}, {Gronoff}, \&
  {Hoekstra}}]{CSW2018b}
{Simon Wedlund}, C., {Behar}, E., {Kallio}, E., {et~al.} 2019{\natexlab{a}},
  accepted in A\&A, 1

\bibitem[{{Simon Wedlund} {et~al.}(2019{\natexlab{b}}){Simon Wedlund},
  {Bodewits}, {Alho}, {Hoekstra}, {Behar}, {Gronoff}, {Gunell}, {Nilsson},
  {Kallio}, \& {Beth}}]{CSW2018a}
{Simon Wedlund}, C., {Bodewits}, D., {Alho}, M., {et~al.} 2019{\natexlab{b}},
  submitted to A\&A, 1

\bibitem[{{Simon Wedlund} {et~al.}(2016){Simon Wedlund}, {Kallio}, {Alho},
  {Nilsson}, {Stenberg Wieser}, {Gunell}, {Behar}, {Pusa}, \&
  {Gronoff}}]{CSW2016}
{Simon Wedlund}, C., {Kallio}, E., {Alho}, M., {et~al.} 2016, A\&A, 587, A154

\bibitem[{{Slavin} \& {Holzer}(1981)}]{Slavin1981}
{Slavin}, J.~A. \& {Holzer}, R.~E. 1981, J. Geophys. Res., 86, 11401

\bibitem[{{Stenberg Wieser} {et~al.}(2017){Stenberg Wieser}, {Odelstad},
  {Wieser}, {Nilsson}, {Goetz}, {Karlsson}, {Andr{\'e}}, {Kalla}, {Eriksson},
  {Nicolaou}, {Simon Wedlund}, {Richter}, \& {Gunell}}]{StenbergWieser2017}
{Stenberg Wieser}, G., {Odelstad}, E., {Wieser}, M., {et~al.} 2017, \mnras,
  469, S522

\bibitem[{{Taylor} {et~al.}(2017){Taylor}, {Altobelli}, {Buratti}, \&
  {Choukroun}}]{Taylor2017}
{Taylor}, M.~G.~G.~T., {Altobelli}, N., {Buratti}, B.~J., \& {Choukroun}, M.
  2017, Phil. Trans. Roy. Soc. London Ser. A, 375, 20160262

\bibitem[{{Uehara} \& {Nikjoo}(2002)}]{Uehara2002}
{Uehara}, S. \& {Nikjoo}, H. 2002, J. Phys. Chem. B, 106, 11051

\bibitem[{{Uehara} {et~al.}(2000){Uehara}, {Toburen}, {Wilson}, {Goodhead}, \&
  {Nikjoo}}]{Uehara2000}
{Uehara}, S., {Toburen}, L.~H., {Wilson}, W.~E., {Goodhead}, D.~T., \&
  {Nikjoo}, H. 2000, Rad. Phys. Chem., 59, 1

\bibitem[{{Volwerk} {et~al.}(2001){Volwerk}, {Kivelson}, \&
  {Khurana}}]{Volwerk2001}
{Volwerk}, M., {Kivelson}, M.~G., \& {Khurana}, K.~K. 2001, \jgr, 106, 26033

\bibitem[{{Volwerk} {et~al.}(2013{\natexlab{a}}){Volwerk}, {Koenders}, {Delva},
  {Richter}, {Schwingenschuh}, {Bentley}, \& {Glassmeier}}]{Volwerk2013b}
{Volwerk}, M., {Koenders}, C., {Delva}, M., {et~al.} 2013{\natexlab{a}}, Ann.
  Geophys., 31, 2213

\bibitem[{{Volwerk} {et~al.}(2013{\natexlab{b}}){Volwerk}, {Koenders}, {Delva},
  {Richter}, {Schwingenschuh}, {Bentley}, \& {Glassmeier}}]{Volwerk2013a}
{Volwerk}, M., {Koenders}, C., {Delva}, M., {et~al.} 2013{\natexlab{b}}, Ann.
  Geophys., 31, 2201

\bibitem[{{Woods} {et~al.}(2005){Woods}, {Eparvier}, {Bailey}, {Chamberlin},
  {Lean}, {Rottman}, {Solomon}, {Tobiska}, \& {Woodraska}}]{Woods2005}
{Woods}, T.~N., {Eparvier}, F.~G., {Bailey}, S.~M., {et~al.} 2005, \jgr (Space
  Physics), 110, A01312

\end{thebibliography}

\end{document}